\documentclass[aps,prb,reprint,superscriptaddress,amsmath,amssymb,floatfix]{revtex4-2}
\usepackage{graphicx,bm,booktabs}
\usepackage{xcolor}
\usepackage[colorlinks=true,citecolor=blue,linkcolor=blue,urlcolor=blue]{hyperref}
\usepackage{CJK}

\newcommand{\kF}{k_{\mathrm{F}}}
\newcommand{\Lam}{\Lambda}
\newcommand{\thr}{\mathrm{thr}}

\begin{document}
\begin{CJK}{UTF8}{ipxm}

\title{\texorpdfstring{Finite-momentum trimers and Cooper quartets in a one-dimensional\\
Fermi gas with coexistent $s$- and $p$-wave interactions}{Finite-momentum trimers and Cooper quartets in a one-dimensional Fermi gas with coexistent s- and p-wave interactions}}

\author{Yixin Guo (郭一昕)}
\email{yixin.guo@riken.jp}
\affiliation{RIKEN Nishina Center for Accelerator-Based Science, Wako 351-0198, Japan}

\date{\today}

\begin{abstract}
We study finite-momentum trimers and Cooper quartets in a one-dimensional two-component Fermi gas with coexistent $s$- and $p$-wave interactions, regularized in the relative momentum of each pair so that they remain Galilean invariant.
Pauli blocking alone then makes the trimer dispersion deviate from free center-of-mass motion, so the trimer energy can be lowest at finite total momentum.
Allowing this motion expands the $abb$ region among pairs and trimers and removes the region where an odd-wave pair is lowest.
The $aabb$ quartet lies below all these branches at every sampled coupling, and in a strip starting on the crossing of the two dimer-pair channels its lowest breakup channel is a moving $abb$ trimer and an atom.
At weak interspecies attraction, a variational bound shows that two odd-wave Cooper pairs bind into a quartet, whose binding exceeds that of the coexisting trimer by orders of magnitude as the attraction weakens.
At stronger interspecies attraction, the quartet binding relative to two unlike-species dimers disappears when the odd-wave interaction is switched off, although the even-wave interaction dominates the interaction energy.
Relative to the in-vacuum case, the Fermi sea enhances the weak-coupling quartet binding, and at fixed scattering lengths the quartet binding grows with density while the coexisting trimer becomes much shallower.
These results show that neither the most strongly bound pair nor the dominant interaction energy alone determines whether the quartet binds.
\end{abstract}

\maketitle

\section{Introduction}\label{sec:intro}

The competition and coexistence of different pairing channels are common issues in superconductivity and fermionic superfluidity.
When several attractive channels are available, their interplay can change both the symmetry and the stability of the resulting state.
In noncentrosymmetric superconductors, antisymmetric spin--orbit coupling allows spin-singlet and spin-triplet components to mix~\cite{Frigeri2004,Smidman2017}.
Mixed even- and odd-parity interband pairing has also been proposed as a mechanism for anapole superconductivity, with UTe$_2$ discussed as a possible candidate~\cite{Kanasugi2022}.
In nuclear systems, the $^1S_0$ and $^3P_2$ pairing channels are relevant to different density regimes of neutron-star matter~\cite{Takatsuka1993}, and a coexistence phase under a magnetic field has been predicted when the coupling between these condensates is included~\cite{Yasui2020}.
These examples show the importance of treating competing attractions together, rather than determining the pairing state from one channel alone.

Attractive fermions can also develop correlations involving more than two particles.
Four-body correlations are particularly relevant in nuclear systems, where the binding of an $\alpha$ particle motivates the study of quartet condensation and its competition with pairing~\cite{Roepke1998}.
Quartet condensation models and quartet Bardeen--Cooper--Schrieffer (BCS) theories have been developed for isovector pairing in self-conjugate nuclei~\cite{Sandulescu2012,Baran2020}, and the quartet BCS description of a quartet superfluid has been compared with a generalized Nambu--Gor'kov formalism~\cite{GuoPRC2026} and applied to the growth of quartet correlations in neutron-rich tellurium isotopes~\cite{GuoPRC2026Te}.
Variational studies have further examined the coexistence of pair and quartet correlations in symmetric nuclear matter~\cite{GuoPRC2022} and the concentration of quartet correlations near the surface of $N=Z$ nuclei~\cite{GuoPRC2025}.
Beyond nuclear physics, biexciton-like Cooper quartets have been investigated in electron--hole liquids~\cite{GuoPRR2022}, while one-dimensional spin-$3/2$ fermions provide a setting for competing pairing and quartetting correlations~\cite{Wu2005}.
Four-body binding can also occur with only two fermionic species: universal $3+1$ tetramers have been predicted in two-dimensional mass-imbalanced mixtures~\cite{Liu2022}, and their relation to a quartet superfluid has been studied~\cite{Liu2023}.
The roles of internal states, mass imbalance, and dimensionality in these systems suggest several possible mechanisms for quartet formation.

Ultracold atomic gases offer control over the interactions and geometry needed to investigate such mechanisms~\cite{Bloch2008,Chin2010}.
The tunability of the $s$-wave attraction has enabled extensive studies of the crossover between BCS pairing and Bose--Einstein condensation of dimers~\cite{Strinati2018}.
Resonant $p$-wave scattering provides access to interactions between identical fermions, as demonstrated in $^{40}$K gases~\cite{Regal2003,Ticknor2004}.
Moreover, neighboring $s$- and $p$-wave Feshbach resonances in $^{40}$K have motivated a proposal for hybridized pairing in a two-component gas~\cite{Zhou2017}.
The coexistence of these interactions also affects few-body binding.
In three dimensions, adding a $p$-wave interaction between identical fermions to their $s$-wave attraction with a third particle can produce shallow trimers, including a Borromean regime~\cite{Naidon2022}.
Thus, the consequences of competing interaction channels extend beyond the symmetry of a pair condensate.

One-dimensional systems are of particular interest because transverse confinement changes scattering and bound-state formation.
Confinement-induced resonances connect the three-dimensional scattering parameters to effective one-dimensional interactions~\cite{Olshanii1998,Bergeman2003}, and confinement-induced dimers have been observed in a Fermi gas~\cite{Moritz2005}.
For $p$-wave interactions, confinement can extend the spatial structure of a shallow molecule~\cite{ZhouCui2017}.
Collisional loss and the confinement dependence of the resonance have also been measured in a quasi-one-dimensional gas~\cite{Chang2020}.
More recently, an emergent even-wave interaction was observed together with odd-wave interactions in a quasi-one-dimensional $^{40}$K gas with active transverse orbital degrees of freedom~\cite{Jackson2023}.
Although these experimental settings have different internal and orbital structures, they motivate models in which even- and odd-wave interactions can both contribute to low-dimensional cluster formation.

The few-body spectrum in one dimension is sensitive to the mass ratio and to interactions between identical particles.
With interspecies attraction, two identical heavy fermions and a light particle can form a trimer when the heavy-to-light mass ratio exceeds unity~\cite{Kartavtsev2009}.
Larger clusters have also been investigated in one-dimensional mass-imbalanced mixtures~\cite{Tononi2022}.
At finite density, studies of asymmetric mixtures have identified many-body regimes with strong trimer correlations and their competition with pairing~\cite{Burovski2009,Orso2010,Dalmonte2012}.
Three-body clustering has also been examined in spinless gases with odd-wave attraction, including the effects of three-body forces~\cite{GuoPRA2022,GuoPRA2023}, and competition among pair and trimer states has been studied in two-component gases with coexistent $s$- and $p$-wave interactions~\cite{GuoPRB2023,GuoPRB2026}.
These results establish the relevance of three-body correlations when assessing whether a larger cluster is energetically favored.

In a degenerate Fermi gas, cluster formation must also be considered in the presence of the occupied Fermi seas.
The Cooper problem shows how an attractive interaction produces a pair above an inert Fermi sea~\cite{Cooper1956}.
Its generalization to three particles has revealed Cooper triples and the effects of Pauli blocking on their spectrum~\cite{Niemann2012}.
Related studies have examined the persistence of three-body correlations in a two-dimensional SU(3) gas~\cite{Kirk2017}, a variational many-body state of Cooper triples~\cite{Akagami2021}, and the crossover between Cooper triples and tightly bound trimers near a triatomic resonance~\cite{Tajima2021}.
Together with the studies of Cooper quartets, these works motivate examining how the Fermi sea affects four-body binding relative to the competing two- and three-body fragments.
This comparison can be addressed within a generalized Cooper calculation before tackling the corresponding many-body state.

For a one-dimensional two-component gas, a quartet containing two fermions of each species offers a different configuration from the four-component and $3+1$ systems discussed above.
In this $aabb$ state, each identical pair can interact in the odd-wave channel, while the four unlike pairs interact in the even-wave channel.
The quartet can dissociate into two unlike-species dimers, two like-species dimers, or a trimer and an atom.
Increasing one attraction therefore changes both the four-body energy and the energies of its fragments.
It is not sufficient to identify the most strongly bound pair: one must determine whether the combined interactions bind the quartet below all allowed breakup channels and how each interaction contributes to that binding.
At finite density, Pauli blocking and the recoil of the fragments further affect this competition.

In this paper, we investigate finite-momentum trimers and the $aabb$ Cooper quartet in a one-dimensional Fermi gas with coexistent $s$- and $p$-wave interactions.
We consider equal densities and use $m_a/m_b=2$ as a representative mass ratio.
Both interactions are regularized in the relative momentum of the interacting pair, so that a moving pair keeps its scattering length.
The pair and trimer comparisons of Refs.~\cite{GuoPRB2023,GuoPRB2026} were made at zero total momentum.
Above the Fermi seas, however, a moving trimer has different Pauli-allowed internal momenta, so its energy is not the zero-momentum value plus a free center-of-mass kinetic energy.
We therefore minimize the trimer energies over their total momentum and determine how trimer motion changes the lowest pair or trimer branch.
Within the same variational approach above inert Fermi seas, we compare these branches with the quartet and determine the quartet binding and its lowest dissociation channel, including trimer--atom channels, in the plane of the two coupling strengths.
We examine the even- and odd-wave interaction contributions within the quartet and construct independent trial states that establish binding at weak and strong interspecies attraction.
We also compare the in-vacuum and in-medium results and follow the trimer and quartet bindings with density at fixed scattering lengths.
These calculations address the formation and stability of individual Cooper clusters.
The cluster-energy ordering provides a starting point for studying the many-body competition among these correlations.

This paper is organized as follows.
The model, variational equations, and breakup thresholds are presented in Sec.~\ref{sec:model}.
Section~\ref{sec:results} discusses weak-coupling quartet formation, the cluster-energy hierarchy with moving trimers, the coupling-plane results, the interaction contributions, the in-vacuum and in-medium comparison, and the density dependence.
A summary is given in Sec.~\ref{sec:summary}.
The Appendixes give the numerical methods and checks, the weak- and strong-coupling trial states, and further trimer branches and dispersions.
We take $\hbar=c=k_{\rm B}=1$ throughout this paper.

\section{Theoretical framework}\label{sec:model}

\subsection{Hamiltonian and variational state}

We consider two fermionic species $a$ and $b$ with masses $m_a$ and $m_b$.
The Hamiltonian is given by
\begin{align}
 H &= K+V_s+V_a+V_b,\\
 K &= \sum_{k,i=a,b}\xi_{k,i}c^\dagger_{k,i}c_{k,i},\\
 V_s &= U_s\sum_{p,p',q}\chi(p)\chi(p')S^\dagger_{p,q}S_{p',q},\\
 V_i &= \frac{U_i}{2}\sum_{p,p',q}\chi(p)\chi(p')\,pp'
 P^\dagger_{i,p,q}P_{i,p',q},
 \qquad i=a,b,
\end{align}
where $\xi_{k,i}=k^2/(2m_i)-\mu_i$, and the pair operators are
\begin{align}
 S^\dagger_{p,q}
 &=c^\dagger_{p+\frac{m_a}{m_a+m_b}q,a}
   c^\dagger_{-p+\frac{m_b}{m_a+m_b}q,b},\\
 P^\dagger_{i,p,q}
 &=c^\dagger_{p+q/2,i}c^\dagger_{-p+q/2,i}.
\end{align}
Here $V_s$ describes even-wave attraction between different species, while $V_a$ and $V_b$ describe odd-wave attraction between identical fermions.
The form factor $\chi(p)=\theta(\Lam-|p|)$ cuts off each interaction in the relative momentum $p$ of the interacting pair.
We assume equal densities, with $k_{{\rm F},a}=k_{{\rm F},b}\equiv\kF$ and $\mu_i=\kF^2/(2m_i)$.
The two components therefore have the same blocked momentum interval but different Fermi energies when $m_a\ne m_b$.

The coupling constants are related to the scattering lengths by
\begin{align}
 U_s&=-\frac{1}{m_r a_s},\qquad m_r=\frac{m_am_b}{m_a+m_b},\\
 \frac{m_i}{2a_p}
 &=\frac{1}{U_i}+\sum_{|p|\le\Lam}\frac{p^2}{2\varepsilon_{p,i}},
 \qquad \varepsilon_{p,i}=\frac{p^2}{2m_i}.
\end{align}
We take $a_s>0$ and a common odd-wave scattering length $a_p$ for the two species.
The corresponding bare couplings $U_a$ and $U_b$ generally differ.
The momentum summation denotes $\int dp/(2\pi)$.
As in Refs.~\cite{GuoPRB2023,GuoPRB2026}, the finite cutoff $\Lam$ specifies the short-range regularization.
Because it acts on relative momenta, the interaction is Galilean invariant: a pair moving with momentum $q$ has the same scattering length as a pair at rest, and an in-vacuum cluster of mass $M$ has the dispersion $E(0)+q^2/(2M)$.
With a cutoff on each single-particle momentum, the allowed relative-momentum interval of an identical pair instead shrinks from $|p|<\Lam$ to $|p|<\Lam-|q|/2$ when the pair moves.
Using the bare coupling fixed at rest then shifts its odd-wave inverse scattering length by $-|q|/\pi$.
Appendix~\ref{app:cutoffshift} derives this shift and shows explicitly why the relative-momentum cutoff used here preserves the scattering length for every $q$.
Few-body quantities retain a dependence on $\Lam$ after fixing the two-body scattering lengths.
We therefore keep $\Lam$ fixed as a model parameter.
Here, numerical grid convergence means that the calculated energies no longer change appreciably when the momentum integrals are discretized more finely, for example by increasing the quadrature order or the number of spectator nodes, with the interaction form factors and $\Lam$ held fixed.
Changing $\Lam$ or the form factor changes the short-range interaction itself and requires a separate physical sensitivity test.

For the quartet at zero total momentum, we introduce
\begin{align}
 |\Psi_4\rangle
 &=\sum_{p_1,p_2,Q}'\Omega_{p_1,p_2,Q}
 c^\dagger_{k_1,a}c^\dagger_{k_2,a}
 c^\dagger_{k_3,b}c^\dagger_{k_4,b}|{\rm FS}\rangle,\\
 k_1&=p_1+Q/2,\qquad k_2=-p_1+Q/2,\nonumber\\
 k_3&=p_2-Q/2,\qquad k_4=-p_2-Q/2.
\end{align}
Hereafter, the prime restricts every constituent momentum to $|k_j|>\kF$, and $|{\rm FS}\rangle$ is the noninteracting Fermi sea.
Fermionic antisymmetry requires $\Omega$ to be odd under either $p_1\to-p_1$ or $p_2\to-p_2$.
The in-vacuum problem is obtained by replacing $|{\rm FS}\rangle$ with $|0\rangle$, removing Pauli blocking, and setting $\mu_i=0$.

Minimizing $\langle\Psi_4|(H-E_4)|\Psi_4\rangle$ gives
\begin{align}
 &(D_{p_1,p_2,Q}-E_4)\Omega_{p_1,p_2,Q}
 +\chi(p_1)p_1\Gamma_a(p_2,Q)\nonumber\\
 &\quad+\chi(p_2)p_2\Gamma_b(p_1,Q)
 +\sum_{\sigma_1,\sigma_2=\pm1}\sigma_1\sigma_2\,\chi(r_{\sigma_1\sigma_2})\nonumber\\
 &\quad\times\Gamma_s\left(-\sigma_1p_1+\frac Q2,-\sigma_2p_2-\frac Q2\right)=0,
 \label{eq:var4}
\end{align}
where $D_{p_1,p_2,Q}=\xi_{k_1,a}+\xi_{k_2,a}+\xi_{k_3,b}+\xi_{k_4,b}$, $r_{\sigma_1\sigma_2}$ is the relative momentum of the interacting $ab$ pair selected by $\sigma_1,\sigma_2$, and
\begin{align}
 \Gamma_a(p_2,Q)&=U_a\sum_{p_1'}'\chi(p_1')p_1'\Omega_{p_1',p_2,Q},\\
 \Gamma_b(p_1,Q)&=U_b\sum_{p_2'}'\chi(p_2')p_2'\Omega_{p_1,p_2',Q},\\
 \Gamma_s(\kappa_a,\kappa_b)
 &=U_s\sum_{Q'}'\chi(r)\,\Omega_{Q'/2-\kappa_a,-Q'/2-\kappa_b,Q'},
\end{align}
with $r$ the relative momentum of the pair summed in $\Gamma_s$.
Solving Eq.~\eqref{eq:var4} for $\Omega$ and inserting it into these definitions yields three coupled equations for amplitudes with two momentum arguments.
We solve them in a Skornyakov--Ter-Martirosyan (STM) form with analytically evaluated pair denominators (Appendixes~\ref{app:grids} and \ref{app:quartet}).
The two-body states and the $aab$ and $abb$ trimers are treated within the same generalized Cooper approach.
The medium enters through the occupation constraints and the Fermi-energy reference.
Particle–hole excitations of the sea are outside this variational space.

\subsection{Trimer parity and breakup thresholds}\label{sec:thresholds}

For an $aab$ trimer with total momentum $P$, let $\psi(k_1,k_2,t)$ be the three-body wave function in the variational state $|\Psi_3\rangle=\sum'_{k_1,k_2}\psi(k_1,k_2,t)\,c^\dagger_{k_1,a}c^\dagger_{k_2,a}c^\dagger_{t,b}|{\rm FS}\rangle$, where $t=P-k_1-k_2$.
Fermionic antisymmetry requires $\psi(k_2,k_1,t)=-\psi(k_1,k_2,t)$.
As for the quartet and as in Refs.~\cite{GuoPRB2023,GuoPRB2026}, minimizing $\langle\Psi_3|(H-E_3)|\Psi_3\rangle$ gives
\begin{align}
 &(D_{k_1,k_2,t}-E_3)\,\psi(k_1,k_2,t)+U_a\chi(p)\,p\,G(t)\nonumber\\
 &\quad+U_s\bigl[\chi(r_{k_1,t})F(k_2)-\chi(r_{k_2,t})F(k_1)\bigr]=0.
 \label{eq:var3}
\end{align}
Here $D_{k_1,k_2,t}=\xi_{k_1,a}+\xi_{k_2,a}+\xi_{t,b}$, $p=(k_1-k_2)/2$ is the relative momentum of the two $a$ atoms, and $r_{k,t}=(m_bk-m_at)/(m_a+m_b)$ is that of an $a$ atom with momentum $k$ and the $b$ atom with momentum $t$.
Because both interactions are separable, the interaction terms depend on $\psi$ only through the pair amplitudes
\begin{align}
 F(\kappa)&=\int'\frac{dk}{2\pi}\chi(r_{k,P-k-\kappa})\,
 \psi(k,\kappa,P-k-\kappa),\\
 G(t)&=\int'\frac{dp}{2\pi}\chi(p)\,p\,
 \psi\left(\frac{P-t}{2}+p,\frac{P-t}{2}-p,t\right).
\end{align}
The amplitude $F(\kappa)$ describes an interacting $ab$ pair with the other $a$ atom as a spectator of momentum $\kappa$, and $G(t)$ an interacting $aa$ pair with the $b$ atom as a spectator of momentum $t$.
Solving Eq.~\eqref{eq:var3} for $\psi$ and inserting it into these definitions gives two coupled integral equations for $F$ and $G$, each a function of a single spectator momentum (Appendix~\ref{app:grids}).
Their lowest root below the breakup threshold introduced below gives the lowest-lying trimer.
At $P=0$, the Hamiltonian is invariant under spatial inversion, which reverses all three momenta, so the trimer states can be classified by their inversion parity $\pi=\pm1$, defined by $\psi(-k_1,-k_2,-t)=\pi\,\psi(k_1,k_2,t)$.
For the amplitudes, this gives
\begin{equation}
 F(-\kappa)=\pi F(\kappa),\qquad G(-t)=-\pi G(t).
 \label{eq:parity}
\end{equation}
Both choices obey fermionic antisymmetry.
We solve both sectors for $aab$ and $abb$.
At nonzero $P$, we use the full momentum axis without a parity restriction.

For an $N$-body cluster ($N=3$ for the trimers and $N=4$ for the quartet) with total momentum $P$, let $E_N(P)$ denote its energy and $E_{\thr,N}(P)$ the lowest energy of the fragments into which it can break up at the same total momentum, that is, its breakup threshold.
The binding energy of the cluster is
\begin{equation}
 B_N(P)=E_{\thr,N}(P)-E_N(P),
 \label{eq:binding}
\end{equation}
and the cluster is bound when $B_N(P)>0$, i.e., when its energy lies below the lowest allowed breakup threshold with the same total momentum.
All in-medium energies are relative to the Fermi sea.
In particular, a positive trimer energy can still describe binding if its breakup threshold is higher.

Denote the $ab$, $aa$, and $bb$ dimer energies by $E_s(Q)$, $E_{paa}(Q)$, and $E_{pbb}(Q)$.
The $aab$ threshold is
\begin{align}
 E_{\thr,aab}(P)=\min\bigl\{&E_{\mathrm{free},aab}(P),\nonumber\\
 &\inf_q'[E_{paa}(P-q)+\xi_{q,b}],\nonumber\\
 &\inf_q'[E_s(P-q)+\xi_{q,a}]\bigr\}.
 \label{eq:thr3}
\end{align}
Here $E_{\mathrm{free},aab}(P)$ is the lowest energy of three unbound particles with total momentum $P$, subject to the Fermi-sea occupation constraints,
\begin{align}
E_{\mathrm{free},aab}(P)
=
\inf'_{\substack{
k_1+k_2+k_3=P
}}
\left(
\xi_{k_1,a}+\xi_{k_2,a}+\xi_{k_3,b}
\right).
\label{eq:free_aab}
\end{align}
The $abb$ expression follows by species interchange.
If a dimer branch does not exist for the parameters considered, the corresponding dimer–particle channel is omitted from the minimization.
The prime denotes minimization over momenta allowed by the Fermi-sea occupation constraints and includes both signs of the allowed spectator momentum.
In the zero-momentum parameter cuts considered below, the atom--dimer minima are at $|q|=\kF$.
For $m_a=2m_b$, $E_{\mathrm{free},3}(0)=3\kF^2/(2m_a)$ for either trimer configuration.

For a zero-momentum quartet, the two-dimer candidate is
\begin{align}
 E_{\mathrm{dd}}=\min_Q\{&2E_s(Q),\,
 E_{paa}(Q)+E_{pbb}(-Q)\}\nonumber\\
 =\min\{&2E_s(0),\,E_{paa}(0)+E_{pbb}(0)\},
 \label{eq:dd}
\end{align}
for the parameters checked here.
The full threshold also includes four atoms, a dimer and two atoms, and
\begin{align}
 E_{aab+b}&=\inf_q'[E_{3,aab}(-q)+\xi_{q,b}],\\
 E_{abb+a}&=\inf_q'[E_{3,abb}(-q)+\xi_{q,a}].
 \label{eq:thr31}
\end{align}
The full trimer dispersion $E_3(P)$, rather than only its zero-momentum value $E_3(0)$, is required in these expressions.
The Pauli-allowed momentum domain is not invariant under a Galilean boost.
Indeed, even the free three-body threshold drops to zero at $P=\kF$.
Since the interaction itself is boost invariant, Pauli blocking is the only source of this momentum dependence.
The quartet binding is always referenced to the lowest checked fragmentation energy.
On the reported density cut, this is $E_{\mathrm{dd}}$.
In part of the coupling plane, it is instead $E_{abb+a}$.
The momentum checks determining this distinction are described in Appendix~\ref{app:breakup}.

\subsection{Parameters and units}

At fixed density we use $m_a=2m_b$ and $\Lam/\kF=10$.
The coupling-plane scan varies both inverse scattering lengths.
As for the representative one-dimensional cut, $1/(\kF a_p)=-1$ is adopted.
Hereafter, we write $g_s=1/(\kF a_s)$ and $g_p=1/(\kF a_p)$.
Momentum and energy units are $\kF$ and $\kF^2/m_b$.
For the density scan, we instead fix $a_s/|a_p|=2$, $a_p<0$, and $\Lam|a_p|=10$.
Its horizontal axis is $\kF|a_p|$ and the energy unit is $1/(m_ba_p^2)$.
Thus, neither the interactions nor the cutoff changes along the density scan.
The calculations concern bound states in the generalized Cooper problem, rather than a thermodynamic phase diagram.

\subsection{Scope of the approximation and short-range sensitivity}
\label{sec:validity}

The generalized Cooper approach describes a few added particles above fixed Fermi seas, retaining Pauli blocking and the Fermi-energy reference.
The interactions within each cluster are treated through the coupled few-body equations.
The approximation concerns the response of the background.
A fixed background is most directly motivated when interactions only weakly modify its occupations and chemical potentials.
Toward stronger $s$-wave attraction and resonant or positive-$g_p$ odd-wave interactions, as in the upper and right parts of Fig.~\ref{fig:hierarchy}, particle--hole dressing and background rearrangement can become important for a quantitative description of the finite-density gas.
Compact clusters with sizes much smaller than $\kF^{-1}$ can nevertheless retain nearly vacuum-like internal binding.
The energy hierarchy and its boundaries characterize the generalized Cooper spectrum at fixed chemical potentials.
Including the background response would allow medium corrections to these boundaries and to the binding relative to competing fragments to be assessed quantitatively.

The cutoff $\Lam$ sets a separate short-range scale.
A low-energy description with reduced sensitivity to the form factor requires both $\kF$ and the characteristic internal relative momenta to be well below $\Lam$.
At $\Lam/\kF=10$, $1/(\Lam a_s)=g_s/10$, so the strong-$s$ cuts reaching $g_s=8$--20 probe distances approaching the short-range scale.
The energies in this regime, including the additional binding of a quartet relative to deeply bound fragments, can therefore be sensitive to the form factor and are presented for the specified finite-cutoff Hamiltonian.
Shallow clusters with internal momenta well below $\Lam$ provide the natural regime for investigating low-energy universality.
Future work can test universality by comparing different form factors while matching the two-body scattering parameters and any relevant few-body short-range observables.
Although background response and short-range corrections may shift the energies and boundary positions, we expect the qualitative mechanisms of trimer and quartet formation arising from Pauli blocking and the interplay of even- and odd-wave attractions to persist.

\section{Results and discussion}\label{sec:results}

The trimer amplitudes depend on a single spectator momentum and are discretized with Gauss--Legendre panels, whose edges include every kink of the amplitudes and are refined toward the Fermi edges (Appendixes~\ref{app:grids} and \ref{app:trimers}).
Trimer results use panels of order 6, that is, six nodes per panel, checked with orders 4 and 8.
The quartet amplitudes depend on two momenta and are solved on a single spectator-momentum grid with $n$ nodes on each side of the Fermi sea, mapped nonlinearly to concentrate near the Fermi momentum (Appendix~\ref{app:quartet}).
Quartet results use $n=18$, checked against $n=14$ and, at selected points, larger $n$.

\subsection{Weak attraction and the binding scales}

We first consider quartet formation at weak interspecies attraction.
At $a_p<0$ there is no odd-wave dimer in vacuum, whereas a finite Fermi momentum produces a nonzero odd-wave form factor at the continuum edge.
The pair equation then contains the Cooper logarithm and supports $aa$ and $bb$ pairs.
A weak $s$-wave attraction acts between these two dimers and can bind them into an $aabb$ quartet.

\begin{figure}[tbp]
\includegraphics[width=\columnwidth]{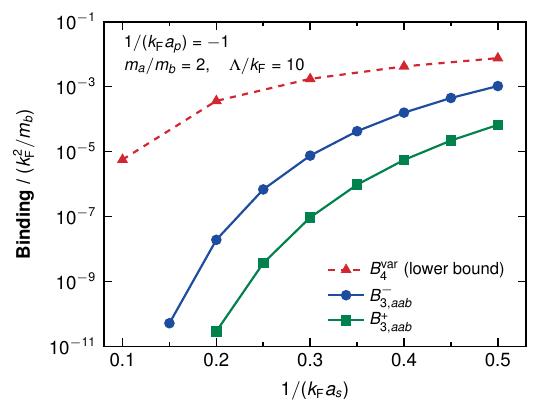}
\caption{\label{fig:weak}
Weak-coupling binding scales at $1/(\kF a_p)=-1$, $m_a/m_b=2$, and $\Lam/\kF=10$.
Circles and squares give the zero-momentum $aab$ trimer bindings $B^-_{3,aab}$ and $B^+_{3,aab}$ in the inversion parities $\pi=-1$ and $+1$, respectively.
Triangles give the variational lower bound $B_4^{\rm var}=E_{\mathrm{dd}}-E_{4,\mathrm{trial}}$ on the quartet binding, where $E_{4,\mathrm{trial}}$ is the energy of the $aa+bb$ trial state.
Each cluster is referenced to its own breakup threshold.
Lines guide the eye between computed points.}
\end{figure}

Figure~\ref{fig:weak} shows this mechanism at $g_p=-1$.
The quartet results use a trial state built from an $aa$ and a $bb$ dimer (Appendix~\ref{app:trial4}), evaluated with the Hamiltonian of Sec.~\ref{sec:model} and independent of the four-body integral-equation discretization.
By the variational principle, its energy expectation value $E_{4,\mathrm{trial}}$ is an upper bound on the lowest quartet energy.
It lies below the two-dimer threshold $E_{\mathrm{dd}}$ throughout the displayed interval, so $B_4^{\rm var}=E_{\mathrm{dd}}-E_{4,\mathrm{trial}}$ is a positive lower bound on the quartet binding.
This bound is $3.64\times10^{-4}$ at $1/(\kF a_s)=0.2$ and $7.54\times10^{-3}$ at 0.5, in units of $\kF^2/m_b$.
The trial state thus establishes a Cooper-type quartet, bound through odd-wave pairs that exist only above the Fermi sea.
It does not require a converged four-body energy root, which the integral-equation grids cannot provide at such small bindings.

The same attraction can bind an $aa$ dimer and a $b$ atom.
Unlike the quartet, the trimer is obtained directly from the three-body integral equations of Sec.~\ref{sec:thresholds}.
The odd-parity $aab$ trimer binding is about $1.04\times10^{-3}$ at $1/(\kF a_s)=0.5$ and falls to $1.9\times10^{-8}$ at 0.2 and $5.2\times10^{-11}$ at 0.15.
The even-parity state is shallower.
Thus, at coupling 0.2, the quartet binding lower bound is already about $4$ orders of magnitude larger than the odd-parity trimer binding.
Both clusters are bound at weak coupling, on binding scales that differ by orders of magnitude, each measured from its own breakup threshold.

The weak quartet binding follows from the attractive projected interaction between the two odd-wave dimers.
As shown in Appendix~\ref{app:weak}, a trial function concentrated near their threshold lowers the energy for arbitrarily weak attraction, as long as the two-dimer channel remains the lowest threshold.
An analogous construction applies to the atom--dimer channel of the trimer.
The rapid decrease of the trimer binding is consistent with this argument.

\subsection{Lowest-lying Cooper cluster}
\label{sec:hierarchy}

We compare single-cluster energies at the fixed chemical potentials $\mu_i=\kF^2/(2m_i)$.
For each pair or trimer configuration $c$, define
\begin{equation}
 \mathcal E_c=\min_P E_c(P),\qquad
 \mathcal E_{\le3}=\min_{c\in\{ab,aa,bb,aab,abb\}}\mathcal E_c.
 \label{eq:hierarchy23}
\end{equation}
The corresponding $P=0$ comparison uses $E_c(0)$ in place of $\mathcal E_c$.
All entries are energies relative to the same Fermi seas, without division by particle number.
This comparison identifies the lowest included Cooper-cluster branch.
It permits higher bound branches to coexist and does not assign many-body populations.

\begin{figure*}[t]
\includegraphics[width=\textwidth]{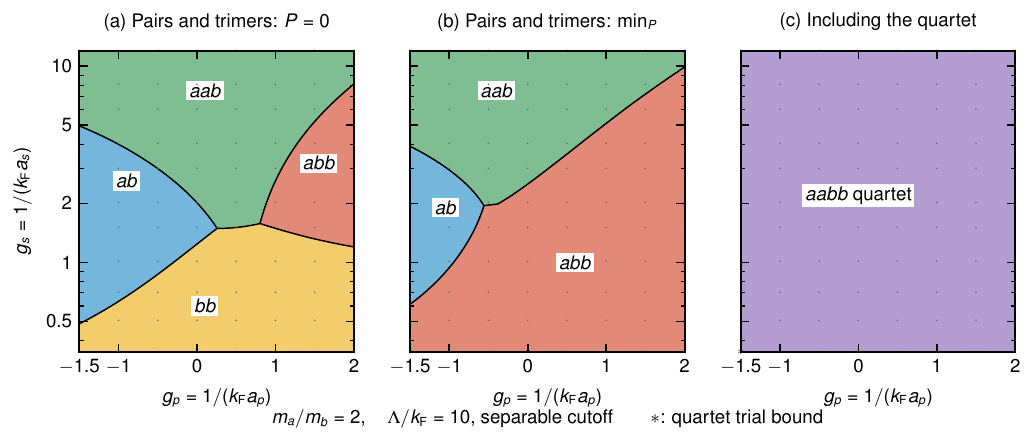}
\caption{\label{fig:hierarchy}
Lowest-lying Cooper-cluster comparison at fixed chemical potentials.
(a) Pairs and trimers restricted to $P=0$.
(b) The comparison after minimizing the trimer energy over its total momentum.
The pair minima are at $P=0$.
(c) Including the quartet, which lies below the pair and trimer branches, minimized over their momenta, at every marked point.
Dots mark integral-equation roots, and the star marks an independent quartet trial bound.
The boundaries in (a) and (b) are solved directly from the STM energy condition (Appendix~\ref{app:trimers}) and joined by straight segments between calculated points.
Markers show the 88 sampled couplings.
Panel (c) has no interior boundary in this window.
All results are for the finite cutoff $\Lam/\kF=10$.}
\end{figure*}

Figure~\ref{fig:hierarchy}(a) retains the zero-momentum comparison used in Refs.~\cite{GuoPRB2023,GuoPRB2026}.
The $bb$ odd-wave pair is lowest at weak $g_s$ and the $ab$ pair at intermediate $g_s$ and negative $g_p$.
Both trimers appear above the junctions $(g_s,g_p)=(1.49,0.26)$ and $(1.58,0.80)$.
In the common window, the boundaries agree with the zero-momentum diagram of Ref.~\cite{GuoPRB2026} within about 0.1 in the couplings.
At stronger interspecies attraction, the $aab$ trimer, rather than the $ab$ pair, becomes the lowest cluster.
The $ab/aab$ boundary leaves the window at $g_s=4.97$ and continues toward more negative $g_p$, reaching $g_p=-9.23$ at $g_s=12$.
This agrees with the vacuum limit, where the odd-parity $aab$ trimer exists for $m_a/m_b>1$ at any coupling~\cite{Kartavtsev2009}.
The $abb$ trimer is lowest at large $g_p$ up to $g_s=8.13$.
Equal odd-wave scattering lengths give $E_{pbb}(0)=2E_{paa}(0)<0$ at $m_a=2m_b$, so the $aa$ pair never has the lowest individual pair energy.

Allowing center-of-mass motion changes this pattern substantially [Fig.~\ref{fig:hierarchy}(b)].
The $abb$ trimer replaces the $bb$ pair and extends to intermediate coupling.
At $(g_s,g_p)=(4,1)$, for example, the zero-momentum $aab$ trimer has energy $-8.14$, below the zero-momentum $abb$ trimer at $-7.61$, but an $abb$ trimer with momentum $2.41\kF$ reaches $-8.61$ and is lower.
The $ab$ region shrinks to $0.61\le g_s\le3.91$ at the left edge of the window.
Below the junction $(g_s,g_p)=(1.95,-0.56)$, the $ab$ region borders the $abb$ trimer at $P=\kF$, determined by the end point condition discussed below.
Above it, the region borders a moving $aab$ trimer whose minimizing momentum grows from $1.31\kF$ to $1.66\kF$ along the boundary.
The $aab/abb$ boundary starts at the same point and leaves the window at $g_s=9.92$, to the left of its zero-momentum counterpart.
Along it, the minimizing momentum of $abb$ grows from $1.5\kF$ to $3.5\kF$, whereas that of $aab$ decreases and vanishes for $g_s\gtrsim7$.
Pauli blocking is the origin of this momentum dependence.
The interaction is Galilean invariant, but moving the cluster changes its allowed internal momenta, which cannot be represented by adding a free center-of-mass kinetic energy to $E_3(0)$.

At weak attraction, the energy gain of the moving trimer can become extremely small.
An $a$ atom near $+\kF$ and a $bb$ dimer near zero momentum provide an $abb$ threshold at $P=\kF$ with energy $E_{pbb}(0)$.
The same edge-localized variational construction used in Appendix~\ref{app:movingabb} lowers this energy for an arbitrarily weak attractive $U_s$.
The $abb$ trimer therefore lies below the $bb$ pair wherever that pair is the lowest, and Fig.~\ref{fig:hierarchy}(b) has no $bb$ region.
At $(g_s,g_p)=(0.5,-1)$, for example, the zero-momentum $bb$ pair has energy $-0.01812$, whereas an $abb$ trimer just above $|P|=\kF$ reaches $-0.0204$.
Where the $ab$ pair is lower, an $ab$ pair at rest and an edge $b$ atom play the same role, and the weak-coupling part of the $ab/abb$ boundary is determined from this $P=\kF$ end point condition.
Finite-offset energy roots approach it only logarithmically (Appendix~\ref{app:trimers}).
The corresponding $aab$ end point lies higher along the whole segment.
The boundary starts at $(g_s,g_p)=(0.61,-1.5)$ and meets the moving $aab$ branch at the junction above.

The quartet lies below the compared pair and moving-trimer branches at every sampled point of Fig.~\ref{fig:hierarchy}(c).
At 77 of the 88 points, no trimer, minimized over its momentum, reaches the two-dimer candidate $E_{\mathrm{dd}}$ of Eq.~\eqref{eq:dd}.
The quartet is bound below $E_{\mathrm{dd}}$ there, and hence below every pair and trimer, without its dispersion.
At the weakest coupling $(g_s,g_p)=(0.35,-1.5)$, this binding is established by the trial state of Appendix~\ref{app:trial4}.
At the remaining 11 points, with $1\le g_s\le4$ and $g_p\ge0$, a moving trimer lies below $E_{\mathrm{dd}}$.
There the lowest quartet root lies below every trimer minimized over its momentum (Appendix~\ref{app:breakup}), for example at $-1.94$ for $(g_s,g_p)=(1,0)$, where $E_{\mathrm{dd}}=-0.41$.
Its advantage over individual pairs is expected when it binds below both dimer-pair channels, since $E_4<2E_s(0)<E_s(0)$ and $E_4<E_{paa}(0)+E_{pbb}(0)<\min[E_{paa}(0),E_{pbb}(0)]$.
Quartet stability against fragments with the same $aabb$ composition requires the separate breakup analysis below.

\begin{figure*}[t]
\includegraphics[width=\textwidth]{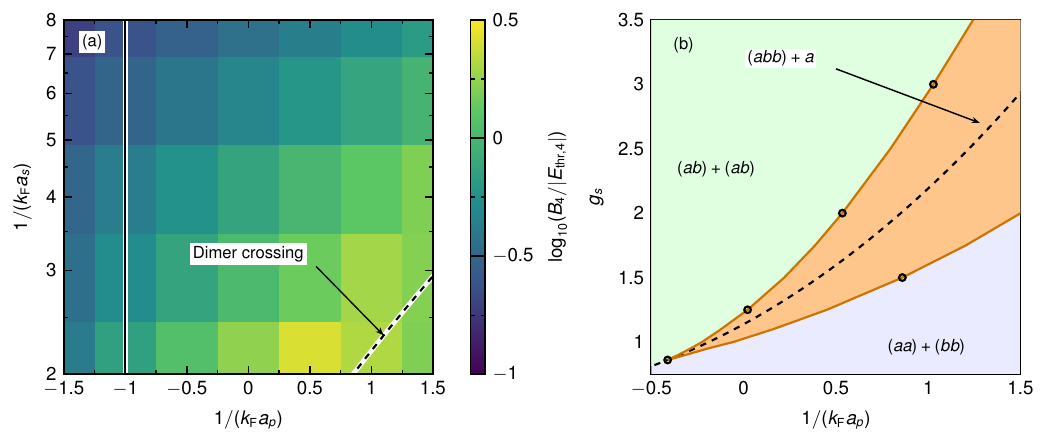}
\caption{\label{fig:quartetmap}
(a) Quartet binding in the coupling plane, with colors indicating $\log_{10}(B_4/|E_{\thr,4}|)$ at 35 sampled couplings in $-1.5\le g_p\le1.5$, $2\le g_s\le8$.
Colors use $n=18$.
Between $n=14$ and 18, $B_4$ changes by at most about 5\%, and $n=22$, a longer axis, or a different node distribution changes it by at most about 3\% at four test points (Appendix~\ref{app:quartet}).
The solid line marks the cut $g_p=-1$ discussed in Sec.~\ref{sec:response}.
(b) The lowest fragmentation channel near the dimer crossing: $(aa)+(bb)$, $(ab)+(ab)$, and $(abb)+a$.
The strip boundaries are solved from moving-trimer calculations.
Circles mark higher-order checks.
The dashed curve in both panels is the equality of the two dimer-pair energies.
Here, $m_a/m_b=2$ and $\Lam/\kF=10$.}
\end{figure*}

\subsection{Quartet binding and dissociation channels in the coupling plane}
\label{sec:quartetmap}

We next consider the $(g_s,g_p)$ coupling plane.
Figure~\ref{fig:quartetmap}(a) shows the quartet binding relative to the magnitude of its lowest breakup threshold $B_4/|E_{\thr,4}|$.
This ratio distinguishes a small energy gain below deeply bound fragments from a quartet whose binding is comparable to the fragment energies.
The dashed curve marks the dimer crossing, where the two dimer-pair channels of Eq.~\eqref{eq:dd} are degenerate $2E_s(0)=E_{paa}(0)+E_{pbb}(0)$.
The $(aa)+(bb)$ channel is lower on its weak-$g_s$ side and $(ab)+(ab)$ on the other.
The displayed window covers $2\le g_s\le8$ and $-1.5\le g_p\le1.5$ at the same mass ratio and cutoff as Fig.~\ref{fig:weak}.
The quartet equations are solved without an inversion-parity restriction, so the lowest root of either sector is obtained.

The quartet is bound at every displayed coupling, by $B_4=0.89$ at $(g_s,g_p)=(2,-1.5)$ up to 22.7 at $(8,1.5)$ in units of $\kF^2/m_b$.
The ratio $B_4/|E_{\thr,4}|$ ranges from 0.20 at $(8,-1.5)$ to 2.6 at $(2,0.5)$.
It grows with the odd-wave attraction at fixed $g_s$.
At fixed $g_p\le0$, the ratio decreases with $g_s$, because the two $ab$ dimers deepen faster than the quartet binding grows.
A moving trimer and an atom provide the threshold at $g_p\ge1$ for $g_s=2$ and at $g_p=1.5$ for $g_s=3$ and 4.
At $g_s=2$ and 3, this lower threshold makes the ratio decrease toward the right edge.

Beyond this window, binding is also established independently of the four-body grids, by variational expectation values at representative weak and strong couplings, built from odd-wave dimers at weak $g_s$ and from antisymmetrized unlike-species dimer profiles at strong $g_s$ (Appendixes~\ref{app:trial4} and \ref{sec:strongtrial}).
There, the aim is only to show that the quartet remains bound, which a variational lower bound does at far lower cost than converging the integral equations on much finer grids.

The actual breakup comparison is richer than the crossing of two pair channels.
Figure~\ref{fig:quartetmap}(b) resolves a strip in which a moving $abb$ trimer and an $a$ atom have the lowest energy.
Its lowest point lies on the dimer crossing, at $(g_s,g_p)=(0.86,-0.41)$.
Along the crossing, the two dimer-pair channels are degenerate, so a third channel undercuts them there first (Appendix~\ref{app:breakup}).
Near this point, the extra $a$ atom sits at the Fermi edge, and the $abb$ trimer at $|P|=\kF$ lies below $E_{pbb}(0)$ by more than $|E_{paa}(0)|$.
The strip broadens toward positive $g_p$ and leaves the plotted range through $g_p=1.5$ at $g_s=2.00$.
The $(aab)+b$ channel also falls below $E_{\mathrm{dd}}$ in part of the strip, starting at $g_s=1.56$ on the crossing, but it stays above $(abb)+a$ at every checked point.
A trimer--atom channel can therefore intervene between the two dimer-pair channels.
The heavier $a$ spectator has a smaller kinetic cost, but the trimer dispersion must also be included to determine the boundary.
A comparison of zero-momentum trimers would miss this region.
Appendix~\ref{sec:trimerchecks} gives further trimer results: the dispersions and trimer--atom energies at $(g_s,g_p)=(0.5,-1)$, outside the strip, where both channels stay above $E_{\mathrm{dd}}$, and the zero-momentum branches of both inversion parities at strong interspecies attraction.

\begin{figure}[tbp]
\includegraphics[width=\columnwidth]{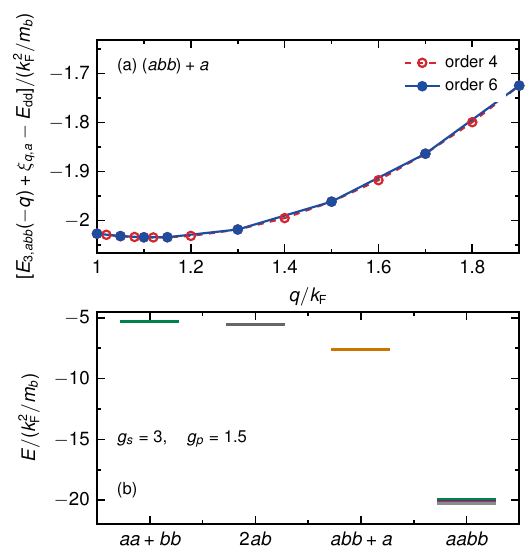}
\caption{\label{fig:channels}
Breakup competition at $g_s=3$ and $g_p=1.5$.
(a) The moving $abb$ trimer plus an $a$ atom, referenced to the lowest two-dimer energy, for moving-trimer panels of order 4 and 6, which coincide on this scale.
(b) The two dimer-pair energies, the minimized trimer--atom energy, and quartet energies on the $n=14,18,22,26$ grids (gray, red, blue, and green lines in the last column).
The quartet lies below all three fragment channels.
Momenta and energies are in units of $\kF$ and $\kF^2/m_b$.
Other parameters are those of Fig.~\ref{fig:quartetmap}.}
\end{figure}

Figure~\ref{fig:channels} gives an example inside the trimer--atom strip.
The minimum occurs at a spectator momentum of approximately $1.12\kF$, and lies about $2.03\kF^2/m_b$ below the two-dimer value.
The quartet lies far below this lower threshold, with $B_4\approx12.3$.
The trimer--atom region in Fig.~\ref{fig:quartetmap}(b) is consequently a change in the preferred fragments, rather than an exclusion region for the quartet.
This distinction is also why a trimer existence map alone does not determine four-body stability.

\begin{figure}[tbp]
\includegraphics[width=\columnwidth]{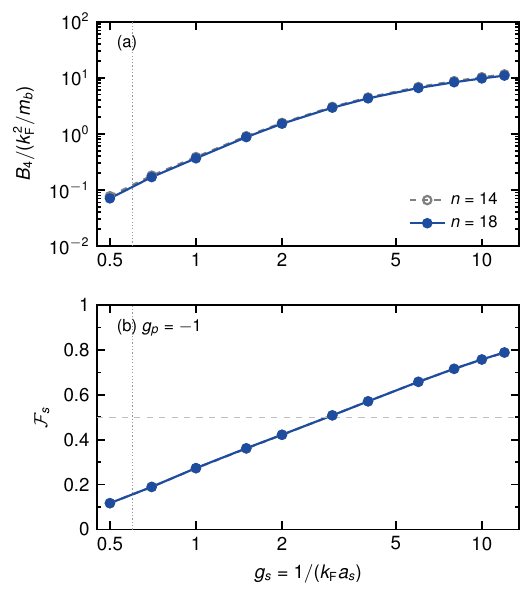}
\caption{\label{fig:quartetcoupling}
Four-body properties along $g_p=-1$.
(a) Quartet binding on the integral-equation grids $n=14$ and 18.
(b) Even-wave fraction $\mathcal F_s=|\langle V_s\rangle|/(|\langle V_s\rangle|+ |\langle V_p\rangle|)$, from energy derivatives on the same grids (Appendix~\ref{app:quartet}).
The vertical dotted line marks the dimer crossing.
The horizontal line in (b) denotes equal even- and odd-wave interaction energies.
Line styles are common to both panels.
Other parameters are those of Fig.~\ref{fig:quartetmap}.}
\end{figure}

\subsection{Binding energy and interaction contributions of the quartet}
\label{sec:response}

Figure~\ref{fig:quartetcoupling}(a) follows the quartet along $g_p=-1$ [solid line in Fig.~\ref{fig:quartetmap}(a)] from $g_s=0.5$, below the dimer crossing, to $g_s=12$.
Its binding grows monotonically with the interspecies attraction, from 0.071 at $g_s=0.5$ to 1.52 at 2 and 10.9 at 12 on the $n=18$ grid.
The relative binding $B_4/|E_{\mathrm{dd}}|$ is largest, about 2.9, near the dimer crossing and decreases to 0.17 at $g_s=12$.
At weaker coupling, binding is established by the trial state of Fig.~\ref{fig:weak}.

The growth of $B_4$ should be distinguished from that of the total quartet energy.
At large $g_s$, two deeply bound $ab$ dimers set the reference energy, and $B_4$ measures the additional gain from joining them.
This gain is carried by the odd-wave interaction: with both odd-wave couplings switched off, the four-body equations have no root below $2E_s(0)$ at $g_s=6$ and 20 (Appendix~\ref{app:quartet}).
The relative-momentum form factors keep the odd-wave attraction of a moving identical pair equal to its value at rest.
A cutoff on single-particle momenta would weaken it (Sec.~\ref{sec:model}), so this binding is sensitive to how the interaction is regularized at a fixed finite cutoff.

To characterize the correlations inside the quartet, we evaluate the interaction energies.
At fixed $\Lam/\kF$, the Hellmann--Feynman relations read
\begin{align}
 \langle V_s\rangle &=g_s\frac{\partial E_4}{\partial g_s},\\
 \langle V_p\rangle &\equiv\langle V_a+V_b\rangle
 =\left(\frac{2\Lam}{\pi\kF}-g_p\right)
   \frac{\partial E_4}{\partial g_p}.
 \label{eq:responses}
\end{align}
Figure~\ref{fig:quartetcoupling}(b) shows the even-wave fraction of the attractive interaction energy.
It increases smoothly from about 0.117 at $g_s=0.5$ to 0.422 at 2 and 0.509 at 3, and reaches 0.789 at 12.
Both grids give the same values.
On this cut, the dimer crossing lies at $g_s=0.60$.
Odd-wave attraction therefore contributes more than half of the interaction energy up to $g_s\approx2.9$, well beyond this dimer crossing, and still about a fifth at $g_s=12$.
These energy fractions are not probabilities of particular dimer configurations.

The role of the odd-wave interaction can also be isolated at stronger coupling, independently of the four-body grid.
We antisymmetrize a product of two $ab$ dimer profiles, as detailed in Appendix~\ref{sec:strongtrial}.
For one such trial state at $(20,-1)$,
\begin{align}
 \langle K\rangle&=36.38167, &\langle V_s\rangle&=-163.18211,\nonumber\\
 \langle V_p\rangle&=-0.87458, &E_{\mathrm{trial}}&=-127.67502.
\end{align}
The checked breakup edge is $E_{\mathrm{dd}}=-127.59606$, giving $B_4\ge0.07896$ within the converged trial integration.
Although $V_p$ contributes only about 0.53\% of this trial state's attractive interaction energy, omitting it from the same expectation value puts the energy above the two-dimer edge.
This comparison shows why dominance of the $s$-wave energy alone does not make the odd-wave channel irrelevant to four-body binding.

\begin{figure}[tbp]
\includegraphics[width=\columnwidth]{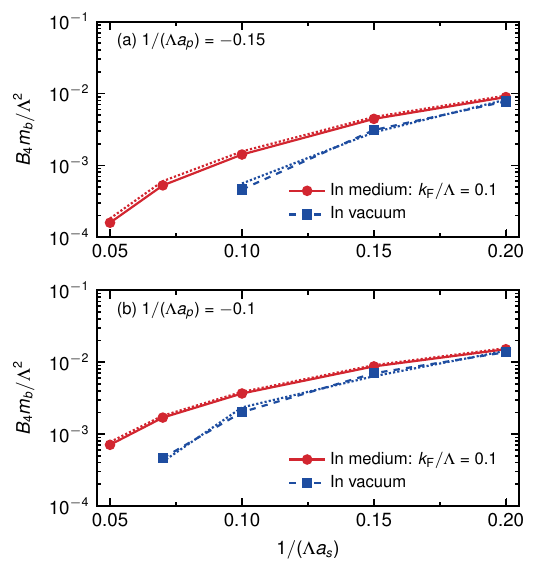}
\caption{\label{fig:vacuumquartet}
Quartet binding in vacuum and in medium at the same scattering lengths and cutoff, for (a) $1/(\Lam a_p)=-0.15$ and (b) $1/(\Lam a_p)=-0.1$.
The in-medium calculation has $\kF/\Lam=0.1$.
Both calculations use $m_a/m_b=2$.
Symbols and connected curves use $n=18$, and dotted curves show $n=14$.
The displayed range starts at $1/(\Lam a_s)=0.05$, below which the grid is not reliable in vacuum (Appendix~\ref{app:quartet}).
All bindings are measured from the appropriate fragmentation threshold.}
\end{figure}

\subsection{In-vacuum and in-medium quartet binding}
\label{sec:vacuum}

To separate the effect of the Fermi sea from a change in interactions, we compare the in-vacuum and in-medium calculations at the same $a_s$, $a_p$, and $\Lam$.
Figure~\ref{fig:vacuumquartet} uses cutoff units, which remain defined at zero density.
At negative $a_p$, the vacuum does not contain an odd-wave dimer, and its lowest fragments are two $ab$ dimers at every plotted coupling.
In medium, the odd-wave Cooper pairs provide the weak-coupling binding mechanism described above.

On the $1/(\Lam a_p)=-0.15$ cut, the in-vacuum binding at $1/(\Lam a_s)=0.1$ is about a third of the in-medium value.
On the $1/(\Lam a_p)=-0.1$ cut, both an energy root and the independent trial state of Fig.~\ref{fig:weak} support an in-medium quartet at $1/(\Lam a_s)=0.05$.
At $1/(\Lam a_s)=0.07$ the in-vacuum binding is about $4.6\times10^{-4}\Lam^2/m_b$ on the $n=18$ grid, compared with $1.7\times10^{-3}\Lam^2/m_b$ in medium.
Toward stronger attraction, the in-vacuum and in-medium bindings approach each other on both cuts.
On the $-0.1$ cut at $1/(\Lam a_s)=0.2$, they are $1.38\times10^{-2}$ and $1.52\times10^{-2}$.
The medium therefore changes the accessible pair channels and the four-body binding scale.

\begin{figure}[tbp]
\includegraphics[width=\columnwidth]{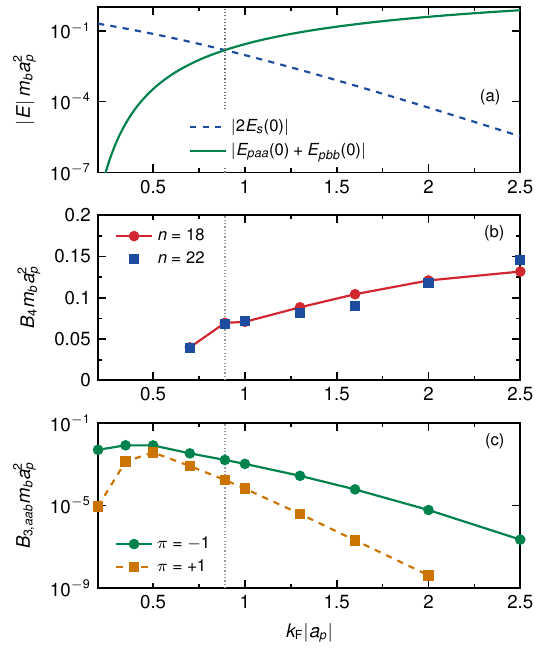}
\caption{\label{fig:density}\label{fig:bindings}
Density dependence at fixed $a_s/|a_p|=2$, $a_p<0$, $\Lam|a_p|=10$, and $m_a/m_b=2$.
(a) Magnitudes of the two negative dimer-pair energies.
The larger magnitude identifies the lower channel.
(b) Quartet bindings on the $n=18$ grid, with $n=22$ checks.
(c) Zero-momentum $aab$ trimer bindings in both inversion parities.
Panels (a) and (c) use logarithmic vertical axes.
The energy unit is $1/(m_ba_p^2)$.
The vertical dotted line marks the dimer crossing at $\kF|a_p|=0.89$.}
\end{figure}

\subsection{Density dependence at fixed interaction}

We next vary $\kF|a_p|$ with $a_s/|a_p|=2$, $a_p<0$, and $\Lam|a_p|=10$ fixed.
At $\kF|a_p|=1$, the scan passes through the point $(g_s,g_p)=(0.5,-1)$ of Fig.~\ref{fig:weak}, with the same $\Lam/\kF=10$.
Figure~\ref{fig:density}(a) shows the two-dimer energies.
The $s$-wave dimers become shallower as the density increases, whereas the odd-wave Cooper pairs become more deeply bound.
The lowest channel changes from $(ab)+(ab)$ to $(aa)+(bb)$ at $\kF|a_p|=0.89$.
This value follows from the pair equations and specifies a change of breakup channel.

The corresponding cluster bindings are shown in Figs.~\ref{fig:density}(b) and \ref{fig:density}(c).
On the $n=18$ grid, the quartet binding increases from 0.040 at $\kF|a_p|=0.7$ to 0.132 at 2.5, and on the $n=22$ grid from 0.039 to 0.145.
It increases monotonically on both grids, by factors of 3.3 and 3.7, whereas the two grids differ by at most 13\% where both are computed.
The slope of $B_4$ drops at the dimer crossing, where the reference threshold changes from $2E_s(0)$ to $E_{paa}(0)+E_{pbb}(0)$.
The plotted values are energy roots below the lowest checked fragmentation threshold.
On this cut, no trimer--atom channel lies below $E_{\mathrm{dd}}$ at the sampled momenta.

The trimer in Fig.~\ref{fig:density}(c) has a different density dependence.
Its odd-parity binding is about $8.2\times10^{-3}$ near $\kF|a_p|=0.35$ and decreases to $2.8\times10^{-4}$ at 1.3, $6.1\times10^{-6}$ at 2, and $2.3\times10^{-7}$ at 2.5.
The even-parity state is shallower.
Thus this density range contains bound trimers together with increasingly stronger quartet binding.

\section{Summary and perspectives}\label{sec:summary}

In this paper, we have investigated finite-momentum trimers and Cooper quartets in a one-dimensional two-component Fermi gas with coexistent $s$- and $p$-wave interactions, both regularized in the relative momentum of the interacting pair.
We have compared pair, trimer, and quartet energies over a broad coupling range at fixed chemical potentials.
Above the Fermi seas, Pauli blocking makes the trimer dispersion deviate from free center-of-mass motion, so the lowest trimer energy can lie at finite momentum.
Along the $aab/abb$ boundary, for example, the minimizing momentum of the $abb$ trimer grows from $1.5\kF$ to $3.5\kF$.
Allowing this motion expands the $abb$ region of the pair and trimer comparison and removes the region where the $bb$ pair is lowest, because a moving $abb$ trimer lies below that pair for arbitrarily weak interspecies attraction.
At strong interspecies attraction, both inversion parities of both trimer configurations are bound, and the odd-parity $aab$ trimer remains the lowest.

The quartet lies below every pair and trimer branch, including moving trimers, at all sampled couplings.
Higher pair and trimer branches can remain bound.
The lowest breakup channel of the quartet is one of the two dimer-pair channels, except in a strip that begins on their crossing and widens toward stronger odd-wave attraction, where a moving $abb$ trimer and an $a$ atom are lowest.
The quartet remains bound at the sampled couplings inside this strip, which therefore marks a change of the preferred fragments rather than an exclusion region.

At weak interspecies attraction, an $aa$ and a $bb$ Cooper pair bind into a quartet through their mutual $s$-wave interaction, which a variational bound establishes independently of the four-body grids.
This quartet is bound far more strongly than the coexisting trimer, by orders of magnitude at the weakest couplings.
As the interspecies attraction increases, the quartet binding grows, and the odd-wave interaction supplies most of its interaction energy well beyond the dimer crossing, before the even-wave share becomes dominant.
At strong interspecies attraction, the binding is nevertheless carried by the odd-wave interaction: without it, the four-body equations have no root below two $ab$ dimers, and in an antisymmetrized two-dimer trial state, an odd-wave share of only 0.53\% of the attractive interaction energy decides binding.
Compared with the in-vacuum case at the same scattering lengths, the Fermi sea supplies odd-wave pairs that do not exist in vacuum at $a_p<0$ and enhances the weak-coupling quartet binding.
Over the density scan at fixed scattering lengths, the quartet binding grows by more than a factor of three, whereas the coexisting trimer becomes shallower by about $4$ orders of magnitude.

The calculations describe a generalized Cooper problem above an inert Fermi sea.
The interplay of even- and odd-wave channels found here is also relevant to mixed-parity superconductors~\cite{Frigeri2004,Smidman2017,Kanasugi2022} and to nuclear systems with coexisting pairing channels and quartet correlations~\cite{Takatsuka1993,Yasui2020,Roepke1998}.
Ultracold gases with neighboring $s$- and $p$-wave resonances and one-dimensional confinement~\cite{Zhou2017,Jackson2023} provide possible settings for testing the cluster-binding mechanisms discussed here.
Particle--hole excitations, self-energy dressing, and cluster lifetimes are not included in the present framework.
Beyond the coupling range of the integral-equation results, independent trial states establish binding at representative points.
The strong-coupling binding also depends on how the interactions are regularized at the finite cutoff.
The applicability of the inert-sea approximation and the role of short-range regularization are discussed in Sec.~\ref{sec:validity}.

Intercluster interactions provide a natural link between the cluster-energy hierarchy and possible many-body phases.
For example, one-dimensional heavy--light $N+1$ clusters can attract and form self-bound structures in the large-$N$ regime studied in Ref.~\cite{Givois2023}, whereas repulsive low-energy trimer--trimer interactions support a stable three-dimensional trimer gas in the parameter regime of Ref.~\cite{Endo2016}.
These studies illustrate how effective intercluster forces can favor a stable cluster gas or self-bound structures in different physical settings.
A quantitative treatment of intercluster interactions and the associated finite-density state lies beyond the scope of the present generalized Cooper calculation.
The cluster spectrum and dissociation channels obtained here provide a basis for studying the many-body competition between three- and four-body correlations.
Future quantitative calculations of trimer--trimer and quartet--quartet scattering will be needed to determine whether the effective intercluster interactions are attractive or repulsive and to assess the stability of the corresponding cluster gas.

\begin{acknowledgments}
The author thanks Lucas Happ, Youngman Kim, Masaaki Kimura, Pascal Naidon, Hiroyuki Sagawa, and Yang Xiao for useful discussions.
Y.G. is supported by the RIKEN Special Postdoctoral Researchers Program.
\end{acknowledgments}

\appendix
\section{Numerical formulation and checks}\label{sec:numerics}

\subsection{Form factors and momentum grids}\label{app:grids}

The pair denominators are the diagonal factors $1+U\Pi$ of the STM equations, whose zeros are the pair bound states.
Here $U$ is the bare coupling of the interacting pair, and $\Pi$ is its pair bubble at the pair's total momentum and at the energy remaining after the spectators' kinetic energies are subtracted.
They are evaluated analytically on their allowed momentum intervals.
The window of a pair is the momentum interval on which its form factor equals one.
Each STM row uses the window of its projected pair as the integration domain and multiplies each exchange term by the window of the pair that term enters.
Constituent momenta are otherwise unlimited.
At $P=0$ and $m_a=2m_b$, the $aab$ amplitudes $F$ and $G$ vanish beyond spectator momenta $3\Lam$ and $1.6\Lam$, respectively, while the $abb$ amplitudes vanish beyond $2.25\Lam$ and $2.5\Lam$.
The spectator axes, which extend to $3\Lam+|P|$, cover these supports.

The amplitudes have kinks where two interval end points cross as the spectator momentum varies.
These end points are the edges of the windows and of the Pauli-blocked intervals.
All of them are linear in the spectator momentum, so the crossing points are computed exactly and used as panel edges.

Two checks test this implementation.
Replacing every window by a single-particle cut reproduces, element by element, the STM matrix of an independent code for that cutoff.
In-vacuum trimers satisfy $E(P)-P^2/(2M)=E(0)$ to all printed digits, as required by Galilean invariance.

\subsection{Pair and trimer calculations}\label{app:trimers}

Shallow zero-momentum trimers need interpolation panels that reach very close to $|k|=\kF$ and $2\kF$, because the continuum edge of the grid is set by the nodes nearest to these points.
The refinement is continued until the plotted bindings no longer change.

For Fig.~\ref{fig:hierarchy}, the coupling at which a pair or trimer reaches a given energy follows from a Schur complement of the STM matrix~\cite{Haynsworth1968, Zhang2005}.
At fixed energy, the odd-wave bare coupling enters the STM matrix only linearly, through the terms acting on the odd-wave amplitude $G$.
Eliminating $F$ therefore turns the condition for a solution into an ordinary eigenvalue problem, whose largest real eigenvalue gives the critical $g_p$.
The critical $g_s$ follows in the same way by eliminating $G$.
For panel (b), this critical coupling is minimized over the total momentum, with a coarse scan of $0\le P\le8\kF$ followed by bracketed minimization.
Since a stronger attraction lowers the trimer energy, this is equivalent to minimizing the trimer energy over $P$.
All minimizing momenta on the plotted boundaries lie below $3.5\kF$, well within the scan.

At $P=\kF$, an $ab$ pair at rest and an atom at the Fermi edge form an atom--dimer threshold with energy $E_s(0)$, the energy against which the trimer is compared.
Near this threshold, the kinetic energy of the edge atom is linear in its distance from the Fermi edge, as in the Cooper problem.
Critical couplings computed at an offset $\delta$ below $E_s(0)$ therefore approach their $\delta\to0$ limit only as $1/\ln(1/\delta)$.
This limit is imposed directly by adding the atom--dimer channel to $F$ as a delta amplitude at the edge.
The Schur complement of the enlarged matrix then gives the zero-offset end point condition, which determines the weak-coupling part of the $ab/abb$ boundary.

For the $aab/abb$ boundary, neither trimer energy is known in advance.
The momentum-minimized critical couplings of the two trimers are therefore computed as functions of a common energy, and the boundary lies where they coincide.
This determines the common energy together with both minimizing momenta.
Where the minimizing $aab$ momentum falls to $P=0$, near $g_s\simeq6$--7, the minimization is repeated over the full momentum scan instead of starting from the previous boundary point.

\subsection{Quartet equations}\label{app:quartet}

For the quartet integral equations, a rational map takes the $n$ Gauss--Legendre nodes per spectator sign onto spectator momenta between $\kF$ and $4\Lam$ in magnitude, and places half of them within about $2\kF$ above $\kF$.
A quartet energy $E_4$ is a root of $\lambda(E)=1$, where $\lambda(E)$ is the largest real eigenvalue of the STM kernel at energy $E$.
Roots are bracketed below the lowest checked fragmentation edge, without an inversion-parity restriction.
The quartet calculations cover the 35 couplings of Fig.~\ref{fig:quartetmap}(a), the 88 binding tests of Fig.~\ref{fig:hierarchy}(c), and the cuts of Figs.~\ref{fig:quartetcoupling}--\ref{fig:density}.

The same axis inserted in the trimer equations reproduces the converged trimer energies to about 1\%, suggesting a relative uncertainty in $B_4$ of order $0.01|E_4|/B_4$.
In the coupling-plane map of Fig.~\ref{fig:quartetmap}(a), comparisons between grid sizes, axis lengths, and node distributions change $B_4$ by a few percent.
Along the density cut of Fig.~\ref{fig:density}, the $n=18$ and 22 bindings differ by up to 13\% where both are computed, while both give the same monotonic increase.
These resolution checks determine the parameter ranges in which quartet bindings are reported.

The grid is also tested for spurious binding of two dimers.
In medium, $\lambda<1$ just below the two-dimer edge with $U_s=0$, where the two odd-wave dimers do not interact.
With both odd-wave couplings switched off ($g_p=-10^3$), $\lambda$ just below $2E_s(0)$ lies between 0.93 and 0.95 at $g_s=6$ and between 0.94 and 0.97 at $g_s=20$ on the $n=14$, 18, and 22 grids for two node distributions, so the even-wave interaction alone gives no root at these couplings.
In vacuum, the grid fails for dimers much larger than $1/\Lam$ and binds even two equal-mass dimers, which the exact solution forbids.
In-vacuum results are therefore shown from $1/(\Lam a_s)=0.05$, where this test is passed.

For the interaction responses, differentiating $\lambda(E_4(x),x)=1$ with respect to a coupling $x$ gives
\begin{equation}
 \frac{\partial E_4}{\partial x}
 =-\frac{\partial\lambda/\partial x}{\partial\lambda/\partial E},
\end{equation}
with both derivatives from symmetric differences of the kernel eigenvalue at the root on the same grid.
They agree with finite differences of the energy roots.

\subsection{Breakup channels and cluster ordering}\label{app:breakup}

The breakup analysis adds the remaining atom's kinetic energy before minimizing over its momentum, and both trimer configurations are retained.
A trimer--atom channel lies below $E_{\mathrm{dd}}$ where some $q\ge\kF$ lets the trimer at momentum $q$ reach $E_{\mathrm{dd}}-\xi_q$.
On the density cut and at the in-medium points of Fig.~\ref{fig:vacuumquartet}, no trimer--atom channel lies below the two-dimer edge at the sampled momenta.

In the coupling plane, this condition defines the strip of Fig.~\ref{fig:quartetmap}(b).
On the $(ab)+(ab)$ side of the dimer crossing, $E_{\mathrm{dd}}$ does not depend on $g_p$, so the critical $g_p$ follows from one Schur complement per $g_s$.
There, increasing $g_p$ lowers the channel relative to $E_{\mathrm{dd}}$.
On the $(aa)+(bb)$ side, $E_{\mathrm{dd}}$ falls faster than the channel as $g_p$ grows.
At fixed $g_s$, the channel is thus lowest relative to $E_{\mathrm{dd}}$ on the crossing.
It undercuts $E_{\mathrm{dd}}$ there first, so the lowest point of the strip is found along the crossing.

The quartet assignments in Fig.~\ref{fig:hierarchy}(c) test binding below the lowest breakup threshold at the 88 displayed couplings.
Where no trimer minimized over its momentum reaches $E_{\mathrm{dd}}$, that is the threshold, and $\lambda(E_{\mathrm{dd}}^-)>1$ on the $n=14$ and 18 grids establishes binding.
Where a moving trimer lies below $E_{\mathrm{dd}}$, the lowest quartet root $E^*$ is determined first.
For each trimer configuration, the Schur complement then gives the smallest $g_p$ at which the trimer reaches $E^*$ at some total momentum.
At all of these points, this value exceeds the actual $g_p$, so no trimer lies below the quartet.
At the weakest coupling, binding is established by the trial state of Appendix~\ref{app:trial4}.
Trial states built from dimers at rest have all constituent momenta below $\Lam$, where every form factor equals one, so their energies do not depend on how the cutoff is imposed.
Together, these tests establish the ordering without identifying the quartet's minimizing momentum.

\section{Variational binding near a fragment threshold}\label{sec:variation}

In this appendix the common identical-particle normalization factors are divided out of both norms and matrix elements.
Angle brackets denote unnormalized matrix elements.
Let $\phi_{i,Q}(k)$ be a normalized odd-wave dimer wave function,
\begin{align}
 \phi_{i,Q}(k)&\propto
 \frac{k-Q/2}{\xi_{k,i}+\xi_{Q-k,i}-E_{pii}(Q)},\\
 \int'\frac{dk}{2\pi}|\phi_{i,Q}(k)|^2&=1.
\end{align}

\subsection{Quartet trial state}\label{app:trial4}

For the quartet, use
\begin{equation}
 \Psi_4=f(Q)\phi_{a,Q}(k_1)\phi_{b,-Q}(k_3).
 \label{eq:trial4}
\end{equation}
The remaining momenta are $k_2=Q-k_1$ and $k_4=-Q-k_3$.
This state is antisymmetric within each species.
With equal $a_p$ and equal blocked intervals, the normalized internal dimer shapes are mass independent.
For real wave functions, define
\begin{equation}
 C(Q,Q')=\int'\frac{d\kappa}{2\pi}
 \phi_{a,Q}(Q-\kappa)\phi_{a,Q'}(Q'-\kappa).
\end{equation}
The integration domain is the intersection of the two allowed domains.
Equivalently, the wave functions are set to zero outside them.
Then,
\begin{align}
 \mathcal N_4&=\int\frac{dQ}{2\pi}|f(Q)|^2,\\
 \langle H_0\rangle_4&=\int\frac{dQ}{2\pi}|f(Q)|^2
 [E_{paa}(Q)+E_{pbb}(-Q)],\\
 \langle V_s\rangle_4&=4U_s\int\frac{dQ\,dQ'}{(2\pi)^2}
 f(Q)f(Q')C(Q,Q')^2,
 \label{eq:trial4V}
\end{align}
where now $H_0=K+V_a+V_b$.
In particular, $C(Q,Q)=1$.
For Fig.~\ref{fig:weak}, we choose $f(Q)=(|Q|+b)^{-1}$ within $|Q|<0.1\kF$ and minimize the Rayleigh quotient over sampled positive $b$.
There, each $\phi_{i,Q}$ is restricted to $|k|,|Q-k|\le\Lam$, and $E_{pii}(Q)$ is replaced by the lowest pair energy on this restricted domain.
All momenta entering an interaction then lie inside $\Lam$, where every form factor equals one, so Eq.~\eqref{eq:trial4V} holds with this replacement.
The restricted energy equals $E_{pii}(0)$ at $Q=0$ and cannot lie below $E_{pii}(Q)$, which keeps the bound valid.
Positive $E_{\mathrm{dd}}-E_{4,\mathrm{trial}}$ gives a lower bound on quartet binding when the two-dimer channel is lowest.

\subsection{Binding at arbitrarily weak attraction}\label{app:weak}

The same kind of trial state describes an $aab$ trimer.
For $aab$ at total momentum zero, choose $\psi=f(q)\phi_{a,-q}(k_1)$, with $k_2=-q-k_1$.
The norm is $\mathcal N_3=\int' dq\,|f(q)|^2/(2\pi)$, and
\begin{align}
 \langle H_0\rangle_3&=\int'\frac{dq}{2\pi}|f(q)|^2
 [E_{paa}(-q)+\xi_{q,b}],\\
 \langle V_s\rangle_3&=2U_s\int'\frac{d\kappa}{2\pi}
 \left|\int'\frac{dq}{2\pi}f(q)\phi_{a,-q}(-q-\kappa)\right|^2,
 \label{eq:trial3}
\end{align}
where $H_0=K+V_a$.
The projected interaction at an edge has a nonzero negative diagonal for $U_s<0$.
Taking $f(-q)=f(q)$ gives total parity $\pi=-1$.
With this choice, both edges $q=\pm\kF$ contribute to Eq.~\eqref{eq:trial3}, and their contributions cannot cancel, because Pauli blocking removes different momenta at the two edges.

The quartet and trimer states have the same local structure.
Let $E_0$ be the fragment energy at its minimum, $E_{\mathrm{dd}}$ for the quartet and the $aa+b$ edge for the trimer.
If $x$ measures the distance from this minimum, the fragment energy exceeds $E_0$ by at most $Cx$ for sufficiently small positive $x$.
For $f(x)=(x+b)^{-1}$, $0<x<\eta$, one obtains
\begin{align}
 \mathcal N&=O(b^{-1}),\qquad L_b=\log(\eta/b),\\
 \langle H_0-E_0\rangle&\le C_1L_b+O(1),\\
 \langle V_s\rangle&\le-C_2|U_s|L_b^2,
 \label{eq:weakproof}
\end{align}
with $C_1,C_2>0$.
For any nonzero attraction the negative term wins at sufficiently small $b$.
This is an existence argument, allowing exponentially small binding, rather than a power-law formula for $B_N$.

It is necessary that $E_0$ be the lowest breakup threshold.
For the weak-$s$ $aab$ branch at $m_a=2m_b$ and $a_p<0$, the odd-wave atom--dimer minimum is at $q=\kF$.
On $\kF\le q\le2\kF$, the allowed relative-momentum interval shrinks and the center-of-mass kinetic energy increases.
For $q\ge2\kF$, the nonnegative in-vacuum odd-wave relative Hamiltonian and the spectator kinetic energy place this channel above the zero-momentum free three-body edge.
The competing $ab+a$ channel is bounded from below by $E_{\mathrm{free},3}-|U_s|\Lam/\pi$.
Its gap above the $aa+b$ edge therefore remains positive at sufficiently weak $U_s$.
For four particles at $U_s=0$, both odd-wave dimers have their lowest energy at zero momentum, and breaking either costs a finite pair binding.
The other fragment channels are consequently separated from $E_{\mathrm{dd}}$.
Since the interaction is bounded at finite cutoff, this separation persists for sufficiently weak $U_s$.
These observations justify Eq.~\eqref{eq:weakproof} relative to the lowest thresholds in a nonzero interval adjacent to $U_s=0$.

\subsection{Moving $abb$ trimer at $P=\kF$}\label{app:movingabb}

For the moving $abb$ branch, choose an $a$ spectator at $q=\kF+x$, $0<x<\eta$, total momentum $P=\kF$, and a $bb$ dimer with momentum $-x$.
The unperturbed energy tends to $E_{pbb}(0)$ as $x\to0$.
Its excess is at most linear in $x$, while the projected $s$-wave interaction has a nonzero attractive diagonal.
The argument of Appendix~\ref{app:weak} therefore also gives an $abb$ trial energy below $E_{pbb}(0)$ for every nonzero attractive $U_s$.
When $bb$ is the lowest pair, this energy is below every three-body fragment threshold, since the extra atom has non-negative energy.
This existence argument does not by itself rank competing trimer branches.

\section{An antisymmetrized two-$ab$-dimer trial state}
\label{sec:strongtrial}

For strong interspecies attraction, a useful independent trial state is built from two unlike-species dimers.
Set all wave functions to zero outside $\kF<|k_i|<\Lam$, and impose $\sum_i k_i=0$, with $k_1,k_2$ for species $a$ and $k_3,k_4$ for $b$.
We choose
\begin{equation}
 \psi=\prod_{i=1}^4u(k_i)
 \left[f(k_1+k_3)-f(k_1+k_4)\right],
 \label{eq:abtrial}
\end{equation}
where $f(Q)=(b+|Q|)^{-1}$ and
\begin{equation}
 u(k)=\left[\xi_{k,a}+\xi_{k,b}-E_s(0)\right]^{-1/2}.
\end{equation}
At opposite internal momenta, $u(k)u(-k)$ has the exact zero-momentum $s$-wave dimer shape.
Equation~\eqref{eq:abtrial} changes sign under exchange of either identical pair and has even inversion parity.
It is an admissible four-body wave function without assuming that its moving internal profiles are dimer eigenstates.
Because every constituent momentum lies inside $\Lam$, each relative momentum does as well, and the form factors equal one on the whole support.

Let $x=\prod_i u(k_i)f(k_1+k_3)$ and let $y$ be its $b$-exchanged partner.
Integrals below include the three independent momentum measures $dk/(2\pi)$ and the original one-particle restrictions.
Exchange symmetry gives
\begin{equation}
 \mathcal N=2\langle x|x-y\rangle,\qquad
 K_0=2\langle x|K|x-y\rangle.
\end{equation}
For example, define the pair contractions
\begin{align}
 F_s(a,b)&=\int'\frac{dk}{2\pi}\,
                  \psi(k,a,-a-b-k,b),\\
 G_a(b_1,b_2)&=\int'\frac{dk}{2\pi}
       \left(k+\frac{b_1+b_2}{2}\right)\nonumber\\
 &\quad\times\psi(k,-b_1-b_2-k,b_1,b_2).
\end{align}
Define $G_b$ analogously by contracting the two $b$ momenta.
The unnormalized interaction matrix elements are
\begin{align}
 V_{s,0}&=4U_s\int'\frac{da\,db}{(2\pi)^2}|F_s(a,b)|^2,\\
 V_{a,0}&=U_a\int'\frac{db_1\,db_2}{(2\pi)^2}|G_a(b_1,b_2)|^2,
\end{align}
with the analogous expression for $V_{b,0}$.
The Rayleigh energy is $(K_0+V_{s,0}+V_{a,0}+V_{b,0})/\mathcal N$.
Identical-particle normalization factors cancel between numerator and denominator.

At $(g_s,g_p)=(20,-1)$, we take $b=0.01\kF$ and evaluate the integrals by composite quadrature that resolves the Pauli edges and the narrow direct and exchange profiles.
Refining this quadrature changes the bound $B_4\ge0.07896$ by less than $10^{-7}$.
The contraction formulas reproduce the expectation value of an independently assembled discrete four-body Hamiltonian on a common test grid, which checks factors and exchange signs.
This trial family also establishes positive binding beyond the quantitative window of Fig.~\ref{fig:quartetmap}(a).
\section{Trimer branches and additional recoil checks}
\label{sec:trimerchecks}

The trimer results in this appendix give additional information about the fragment spectrum.

\subsection{Trimer branches at stronger attraction}

\begin{figure}[tbp]
\includegraphics[width=\columnwidth]{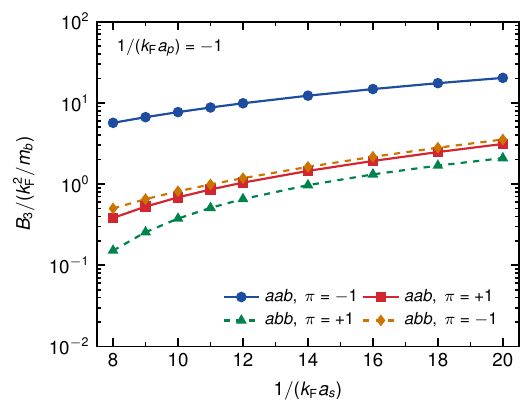}
\caption{\label{fig:parities}
Zero-momentum trimer branches at stronger $s$-wave attraction, with the same mass ratio, odd-wave coupling, and cutoff as Fig.~\ref{fig:weak}.
All four configuration and parity sectors are bound.
The vertical axis is logarithmic.
Lines join calculated points.}
\end{figure}

Both inversion parities $\pi=\pm1$ of both configurations support zero-momentum trimers on the strong-coupling cut, as shown in Fig.~\ref{fig:parities}.
Their common breakup threshold is an $ab$ pair moving with momentum $\kF$ together with an atom at the Fermi edge.
A pair at rest cannot be used, because it cannot combine with an allowed atom to zero total momentum.
All four branches deepen as $g_s$ increases.
The odd-parity $aab$ branch is by far the deepest, with $B_3=5.68$ at $1/(\kF a_s)=8$ and 20.3 at 20, and it remains the lowest trimer throughout.
It has the parity of the heavy--heavy--light trimer that exists in vacuum for $m_a/m_b>1$~\cite{Kartavtsev2009}, and at these couplings it lies below the $ab$ pair at rest, as in Fig.~\ref{fig:hierarchy}(a).

The even-parity branches and the odd-parity $abb$ branch are shallower, with bindings of order one: at $1/(\kF a_s)=16$, for example, 1.93 for the even $aab$ branch, 2.17 and 1.32 for the odd and even $abb$ branches.
The four branches are labeled by both the exchange symmetry and the inversion parity.

\subsection{Moving fragments and quartet stability}

\begin{figure}[tbp]
\includegraphics[width=\columnwidth]{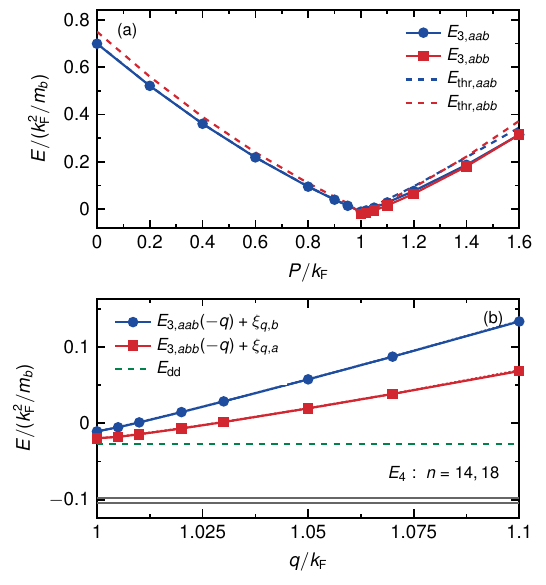}
\caption{\label{fig:dispersion}
(a) Moving-trimer energies and atom--dimer thresholds at $1/(\kF a_s)=0.5$, $1/(\kF a_p)=-1$, $m_a/m_b=2$, and $\Lam/\kF=10$.
A trimer curve is shown only where a root lies below its threshold.
For $P\le\kF$, the $aab$ threshold lies less than $2\times10^{-3}$ above the trimer and is hidden behind its curve.
(b) Trimer--atom energies relevant to a zero-momentum quartet, enlarged near the smallest allowed spectator momentum.
Solid curves with symbols use moving-trimer panels of order 6.
Dotted curves in (b) use order 4.
The dashed horizontal line is $E_{\mathrm{dd}}$.
The two lower lines are quartet energies at $n=14$ and 18, indicating their resolution dependence.}
\end{figure}

Figure~\ref{fig:dispersion}(a) shows the trimer dispersions at $1/(\kF a_s)=0.5$.
Their momentum dependence is substantial.
The $aab$ energy decreases from approximately $0.699$ at $P=0$ to $-0.0106$ at $P=\kF$.
An $abb$ trimer exists from $P=\kF$ upward, and no $abb$ root lies below the threshold at smaller momenta.
The changes in the atom--dimer thresholds follow from the redistribution of momentum among the constituents, with each momentum still outside the Fermi sea.
The trimer bindings themselves stay between $4\times10^{-4}$ and $6\times10^{-2}$ over the displayed momenta.

The spectator cost must be added before using either moving trimer in a four-body breakup channel.
Figure~\ref{fig:dispersion}(b) shows these energies near $q=\kF$.
At that edge, the $aab+b$ and $abb+a$ values are approximately $-0.0106$ and $-0.0196$, respectively, compared with $E_{\mathrm{dd}}=-0.02719$.
Both rise with $q$, and since no trimer minimized over its momentum reaches $E_{\mathrm{dd}}$ at this coupling, neither channel falls below it.
The quartet energy is lower still, at $E_4=-0.0980$ on the $n=18$ grid.
The two-dimer comparison is therefore consistent with the explicit moving-fragment check at this point.
The finite momentum of the fragments is essential to this check, even when it does not change the lowest channel.

\section{Scattering-length shift from a single-particle cutoff}
\label{app:cutoffshift}

For two identical fermions of species $i$, write their momenta as $q/2\pm p$ and measure the energy $z$ from the free center-of-mass energy $q^2/(4m_i)$.
Imposing $|q/2\pm p|<\Lam$ gives, for $|q|<2\Lam$,
\begin{equation}
 |p|<L_q,\qquad L_q=\Lam-\frac{|q|}{2}.
 \label{eq:cutoffwindow}
\end{equation}
The scalar part of the inverse odd-wave $T$ matrix is
\begin{equation}
 \tau_q^{-1}(z)=\frac{1}{U_i}
 -\int_{-L_q}^{L_q}\frac{dp}{2\pi}
       \frac{p^2}{z-p^2/m_i+i0}.
 \label{eq:cutofftmatrix}
\end{equation}
The zero-energy limit of the integral is $-m_iL_q/\pi$.
The bare coupling fixed at rest obeys $U_i^{-1}=m_i/(2a_p)-m_i\Lam/\pi$, so
\begin{equation}
\begin{aligned}
 \tau_q^{-1}(0)&=\frac{m_i}{2a_p}
       -\frac{m_i\Lam}{\pi}+\frac{m_iL_q}{\pi}\\
       &=\frac{m_i}{2}\left(\frac{1}{a_p}-\frac{|q|}{\pi}\right).
 \end{aligned}
 \label{eq:cutoffinverse}
\end{equation}
Defining the effective inverse scattering length by the same zero-energy convention $\tau_q^{-1}(0)=m_i/[2a_p^{\mathrm{eff}}(q)]$ gives
\begin{equation}
\frac{1}{a_p^{\mathrm{eff}}(q)}
       =\frac{1}{a_p}-\frac{|q|}{\pi}.
 \label{eq:cutoffshift}
\end{equation}
The bubble diverges linearly because its integrand approaches a nonzero constant at large $|p|$.
Removing an interval of total width $|q|$ therefore leaves a finite shift even as $\Lam$ increases at fixed $q$.
With the relative-momentum form factor used in this paper, the integration bounds are instead $-\Lam$ and $\Lam$ for every $q$.
The cutoff contribution $m_i\Lam/\pi$ to $\tau_q^{-1}(0)$ then cancels the $-m_i\Lam/\pi$ term in $U_i^{-1}$, giving $\tau_q^{-1}(0)=m_i/(2a_p)$.
Thus, $[a_p^{\mathrm{eff}}(q)]^{-1}=a_p^{-1}$ for every total momentum $q$.

\end{CJK}

\begin{thebibliography}{99}

\bibitem{Frigeri2004} P.~A.~Frigeri, D.~F.~Agterberg, A.~Koga, and M.~Sigrist,
  Superconductivity without inversion symmetry: MnSi versus CePt$_3$Si,
  \href{https://doi.org/10.1103/PhysRevLett.92.097001}{Phys.\ Rev.\ Lett.\ \textbf{92}, 097001 (2004)}.

\bibitem{Smidman2017} M.~Smidman, M.~B.~Salamon, H.~Q.~Yuan, and D.~F.~Agterberg,
  Superconductivity and spin--orbit coupling in non-centrosymmetric materials: A review,
  \href{https://doi.org/10.1088/1361-6633/80/3/036501}{Rep.\ Prog.\ Phys.\ \textbf{80}, 036501 (2017)}.

\bibitem{Kanasugi2022} S.~Kanasugi and Y.~Yanase,
  Anapole superconductivity from $\mathcal{PT}$-symmetric mixed-parity interband pairing,
  \href{https://doi.org/10.1038/s42005-022-00804-7}{Commun.\ Phys.\ \textbf{5}, 39 (2022)}.

\bibitem{Takatsuka1993} T.~Takatsuka and R.~Tamagaki,
  Superfluidity in neutron star matter and symmetric nuclear matter,
  \href{https://doi.org/10.1143/PTPS.112.27}{Prog.\ Theor.\ Phys.\ Suppl.\ \textbf{112}, 27 (1993)}.

\bibitem{Yasui2020} S.~Yasui, D.~Inotani, and M.~Nitta,
  Coexistence phase of ${}^1S_0$ and ${}^3P_2$ superfluids in neutron stars,
  \href{https://doi.org/10.1103/PhysRevC.101.055806}{Phys.\ Rev.\ C\ \textbf{101}, 055806 (2020)}.

\bibitem{Roepke1998} G.~R\"opke, A.~Schnell, P.~Schuck, and P.~Nozi\`eres,
  Four-particle condensate in strongly coupled fermion systems,
  \href{https://doi.org/10.1103/PhysRevLett.80.3177}{Phys.\ Rev.\ Lett.\ \textbf{80}, 3177 (1998)}.

\bibitem{Sandulescu2012} N.~Sandulescu, D.~Negrea, J.~Dukelsky, and C.~W.~Johnson,
  Quartet condensation and isovector pairing correlations in $N=Z$ nuclei,
  \href{https://doi.org/10.1103/PhysRevC.85.061303}{Phys.\ Rev.\ C\ \textbf{85}, 061303(R) (2012)}.

\bibitem{Baran2020} V.~V.~Baran and D.~S.~Delion,
  A quartet BCS-like theory,
  \href{https://doi.org/10.1016/j.physletb.2020.135462}{Phys.\ Lett.\ B\ \textbf{805}, 135462 (2020)}.

\bibitem{GuoPRC2026} Y.~Guo, H.~Tajima, and H.~Liang,
  Comparative study of a quartet superfluid state: Quartet Bardeen--Cooper--Schrieffer theory and generalized Nambu--Gor'kov formalism,
  \href{https://doi.org/10.1103/snvw-f54b}{Phys.\ Rev.\ C\ \textbf{114}, 034304 (2026)}.

\bibitem{GuoPRC2026Te} Y.~Guo, H.~Sagawa, and M.~Kimura,
  Growth of quartet correlations in neutron-rich tellurium isotopes within quartet Bardeen--Cooper--Schrieffer theory,
  \href{https://doi.org/10.1103/5hsp-ccsn}{Phys.\ Rev.\ C\ \textbf{114}, 034329 (2026)}.

\bibitem{GuoPRC2022} Y.~Guo, H.~Tajima, and H.~Liang,
  Cooper quartet correlations in infinite symmetric nuclear matter,
  \href{https://doi.org/10.1103/PhysRevC.105.024317}{Phys.\ Rev.\ C\ \textbf{105}, 024317 (2022)}.

\bibitem{GuoPRC2025} Y.~Guo, T.~Naito, H.~Tajima, and H.~Liang,
  Quartet correlations near the surface of $N=Z$ nuclei,
  \href{https://doi.org/10.1103/4rqf-5kfx}{Phys.\ Rev.\ C\ \textbf{112}, 024310 (2025)}.

\bibitem{GuoPRR2022} Y.~Guo, H.~Tajima, and H.~Liang,
  Biexciton-like quartet condensates in an electron--hole liquid,
  \href{https://doi.org/10.1103/PhysRevResearch.4.023152}{Phys.\ Rev.\ Res.\ \textbf{4}, 023152 (2022)}.

\bibitem{Wu2005} C.~Wu,
  Competing orders in one-dimensional spin-$3/2$ fermionic systems,
  \href{https://doi.org/10.1103/PhysRevLett.95.266404}{Phys.\ Rev.\ Lett.\ \textbf{95}, 266404 (2005)}.

\bibitem{Liu2022} R.~Liu, C.~Peng, and X.~Cui,
  Universal tetramer and pentamer bound states in two-dimensional fermionic mixtures,
  \href{https://doi.org/10.1103/PhysRevLett.129.073401}{Phys.\ Rev.\ Lett.\ \textbf{129}, 073401 (2022)}.

\bibitem{Liu2023} R.~Liu, W.~Wang, and X.~Cui,
  Quartet superfluid in two-dimensional mass-imbalanced Fermi mixtures,
  \href{https://doi.org/10.1103/PhysRevLett.131.193401}{Phys.\ Rev.\ Lett.\ \textbf{131}, 193401 (2023)}.

\bibitem{Bloch2008} I.~Bloch, J.~Dalibard, and W.~Zwerger,
  Many-body physics with ultracold gases,
  \href{https://doi.org/10.1103/RevModPhys.80.885}{Rev.\ Mod.\ Phys.\ \textbf{80}, 885 (2008)}.

\bibitem{Chin2010} C.~Chin, R.~Grimm, P.~Julienne, and E.~Tiesinga,
  Feshbach resonances in ultracold gases,
  \href{https://doi.org/10.1103/RevModPhys.82.1225}{Rev.\ Mod.\ Phys.\ \textbf{82}, 1225 (2010)}.

\bibitem{Strinati2018} G.~C.~Strinati, P.~Pieri, G.~R\"opke, P.~Schuck, and M.~Urban,
  The BCS--BEC crossover: From ultra-cold Fermi gases to nuclear systems,
  \href{https://doi.org/10.1016/j.physrep.2018.02.004}{Phys.\ Rep.\ \textbf{738}, 1 (2018)}.

\bibitem{Regal2003} C.~A.~Regal, C.~Ticknor, J.~L.~Bohn, and D.~S.~Jin,
  Tuning $p$-wave interactions in an ultracold Fermi gas of atoms,
  \href{https://doi.org/10.1103/PhysRevLett.90.053201}{Phys.\ Rev.\ Lett.\ \textbf{90}, 053201 (2003)}.

\bibitem{Ticknor2004} C.~Ticknor, C.~A.~Regal, D.~S.~Jin, and J.~L.~Bohn,
  Multiplet structure of Feshbach resonances in nonzero partial waves,
  \href{https://doi.org/10.1103/PhysRevA.69.042712}{Phys.\ Rev.\ A\ \textbf{69}, 042712 (2004)}.

\bibitem{Zhou2017} L.~Zhou, W.~Yi, and X.~Cui,
  Fermion superfluid with hybridized $s$- and $p$-wave pairings,
  \href{https://doi.org/10.1007/s11433-017-9087-7}{Sci.\ China Phys.\ Mech.\ Astron.\ \textbf{60}, 127011 (2017)}.

\bibitem{Naidon2022} P.~Naidon, L.~Pricoupenko, and C.~Schmickler,
  Shallow trimers of two identical fermions and one particle in resonant regimes,
  \href{https://doi.org/10.21468/SciPostPhys.12.6.185}{SciPost Phys.\ \textbf{12}, 185 (2022)}.

\bibitem{Olshanii1998} M.~Olshanii,
  Atomic scattering in the presence of an external confinement and a gas of impenetrable bosons,
  \href{https://doi.org/10.1103/PhysRevLett.81.938}{Phys.\ Rev.\ Lett.\ \textbf{81}, 938 (1998)}.

\bibitem{Bergeman2003} T.~Bergeman, M.~G.~Moore, and M.~Olshanii,
  Atom--atom scattering under cylindrical harmonic confinement: Numerical and analytic studies of the confinement induced resonance,
  \href{https://doi.org/10.1103/PhysRevLett.91.163201}{Phys.\ Rev.\ Lett.\ \textbf{91}, 163201 (2003)}.

\bibitem{Moritz2005} H.~Moritz, T.~St\"oferle, K.~G\"unter, M.~K\"ohl, and T.~Esslinger,
  Confinement induced molecules in a 1D Fermi gas,
  \href{https://doi.org/10.1103/PhysRevLett.94.210401}{Phys.\ Rev.\ Lett.\ \textbf{94}, 210401 (2005)}.

\bibitem{ZhouCui2017} L.~Zhou and X.~Cui,
  Stretching $p$-wave molecules by transverse confinements,
  \href{https://doi.org/10.1103/PhysRevA.96.030701}{Phys.\ Rev.\ A\ \textbf{96}, 030701(R) (2017)}.

\bibitem{Chang2020} Y.-T.~Chang, R.~Senaratne, D.~Cavazos-Cavazos, and R.~G.~Hulet,
  Collisional loss of one-dimensional fermions near a $p$-wave Feshbach resonance,
  \href{https://doi.org/10.1103/PhysRevLett.125.263402}{Phys.\ Rev.\ Lett.\ \textbf{125}, 263402 (2020)}.

\bibitem{Jackson2023} K.~G.~Jackson, C.~J.~Dale, J.~Maki, K.~G.~S.~Xie, B.~A.~Olsen, D.~J.~M.~Ahmed-Braun, S.~Zhang, and J.~H.~Thywissen,
  Emergent $s$-wave interactions between identical fermions in quasi-one-dimensional geometries,
  \href{https://doi.org/10.1103/PhysRevX.13.021013}{Phys.\ Rev.\ X\ \textbf{13}, 021013 (2023)}.

\bibitem{Kartavtsev2009} O.~I.~Kartavtsev, A.~V.~Malykh, and S.~A.~Sofianos,
  Bound states and scattering lengths of three two-component particles with zero-range interactions under one-dimensional confinement,
  \href{https://doi.org/10.1134/S1063776109030017}{J.\ Exp.\ Theor.\ Phys.\ \textbf{108}, 365 (2009)}.

\bibitem{Tononi2022} A.~Tononi, J.~Givois, and D.~S.~Petrov,
  Binding of heavy fermions by a single light atom in one dimension,
  \href{https://doi.org/10.1103/PhysRevA.106.L011302}{Phys.\ Rev.\ A\ \textbf{106}, L011302 (2022)}.

\bibitem{Burovski2009} E.~Burovski, G.~Orso, and T.~Jolicoeur,
  Multiparticle composites in density-imbalanced quantum fluids,
  \href{https://doi.org/10.1103/PhysRevLett.103.215301}{Phys.\ Rev.\ Lett.\ \textbf{103}, 215301 (2009)}.

\bibitem{Orso2010} G.~Orso, E.~Burovski, and T.~Jolicoeur,
  Luttinger liquid of trimers in Fermi gases with unequal masses,
  \href{https://doi.org/10.1103/PhysRevLett.104.065301}{Phys.\ Rev.\ Lett.\ \textbf{104}, 065301 (2010)}.

\bibitem{Dalmonte2012} M.~Dalmonte, K.~Dieckmann, T.~Roscilde, C.~Hartl, A.~E.~Feiguin, U.~Schollw\"ock, and F.~Heidrich-Meisner,
  Dimer, trimer, and Fulde--Ferrell--Larkin--Ovchinnikov liquids in mass- and spin-imbalanced trapped binary mixtures in one dimension,
  \href{https://doi.org/10.1103/PhysRevA.85.063608}{Phys.\ Rev.\ A\ \textbf{85}, 063608 (2012)}.

\bibitem{GuoPRA2022} Y.~Guo and H.~Tajima,
  Stability against three-body clustering in one-dimensional spinless $p$-wave fermions,
  \href{https://doi.org/10.1103/PhysRevA.106.043310}{Phys.\ Rev.\ A\ \textbf{106}, 043310 (2022)}.

\bibitem{GuoPRA2023} Y.~Guo and H.~Tajima,
  Cooper pairing and tripling in one-dimensional spinless fermions with attractive two- and three-body forces,
  \href{https://doi.org/10.1103/PhysRevA.108.043303}{Phys.\ Rev.\ A\ \textbf{108}, 043303 (2023)}.

\bibitem{GuoPRB2023} Y.~Guo and H.~Tajima,
  Competition between pairing and tripling in one-dimensional fermions with coexistent $s$- and $p$-wave interactions,
  \href{https://doi.org/10.1103/PhysRevB.107.024511}{Phys.\ Rev.\ B\ \textbf{107}, 024511 (2023)}.

\bibitem{GuoPRB2026} Y.~Guo,
  Mass-imbalance effect on the cluster formation in a one-dimensional Fermi gas with coexistent $s$- and $p$-wave interactions,
  \href{https://doi.org/10.1103/qhf4-yr35}{Phys.\ Rev.\ B\ \textbf{113}, 054512 (2026)}.

\bibitem{Cooper1956} L.~N.~Cooper,
  Bound electron pairs in a degenerate Fermi gas,
  \href{https://doi.org/10.1103/PhysRev.104.1189}{Phys.\ Rev.\ \textbf{104}, 1189 (1956)}.

\bibitem{Niemann2012} P.~Niemann and H.-W.~Hammer,
  Pauli-blocking effects and Cooper triples in three-component Fermi gases,
  \href{https://doi.org/10.1103/PhysRevA.86.013628}{Phys.\ Rev.\ A\ \textbf{86}, 013628 (2012)}.

\bibitem{Kirk2017} T.~Kirk and M.~M.~Parish,
  Three-body correlations in a two-dimensional SU(3) Fermi gas,
  \href{https://doi.org/10.1103/PhysRevA.96.053614}{Phys.\ Rev.\ A\ \textbf{96}, 053614 (2017)}.

\bibitem{Akagami2021} S.~Akagami, H.~Tajima, and K.~Iida,
  Condensation of Cooper triples,
  \href{https://doi.org/10.1103/PhysRevA.104.L041302}{Phys.\ Rev.\ A\ \textbf{104}, L041302 (2021)}.

\bibitem{Tajima2021} H.~Tajima, S.~Tsutsui, T.~M.~Doi, and K.~Iida,
  Three-body crossover from a Cooper triple to a bound trimer state in three-component Fermi gases near a triatomic resonance,
  \href{https://doi.org/10.1103/PhysRevA.104.053328}{Phys.\ Rev.\ A\ \textbf{104}, 053328 (2021)}.
  
  \bibitem{Givois2023} J.~Givois, A.~Tononi, and D.~S.~Petrov,
  Self-binding of one-dimensional fermionic mixtures with zero-range interspecies attraction,
  \href{https://doi.org/10.21468/SciPostPhys.14.5.091}{SciPost Phys.\ \textbf{14}, 091 (2023)}.

\bibitem{Endo2016} S.~Endo, A.~M.~Garc\'{i}a-Garc\'{i}a, and P.~Naidon,
  Universal clusters as building blocks of stable quantum matter,
  \href{https://doi.org/10.1103/PhysRevA.93.053611}{Phys.\ Rev.\ A\ \textbf{93}, 053611 (2016)}.
  
  \bibitem{Haynsworth1968} E.~V.~Haynsworth,
Determination of the inertia of a partitioned Hermitian matrix,
\href{https://doi.org/10.1016/0024-3795(68)90050-5} {Linear Algebra Appl.\ \textbf{1}, 73 (1968)}.

 \bibitem{Zhang2005} Edited by F.~Zhang,
\textit{The Schur Complement and Its Applications}, Numerical Methods and Algorithms Vol.~4 (Springer, New York, 2005),
\href{https://doi.org/10.1007/b105056}{doi:10.1007/b105056}.


\end{thebibliography}
\end{document}